\documentclass[12pt]{article}

\usepackage[pdftex]{graphicx}
\usepackage{times}

\newenvironment{sciabstract}{%
\begin{quote} \bf}
{\end{quote}}

\title{Tuning quantum transport by controlling spin reorientations in Dirac semimetal candidates Eu$_{1-x}$Sr$_{x}$MnSb$_{2}$}

\author
{Qiang Zhang$^{1,2\ast+}$, Jinyu Liu$^{3+}$, Huibo Cao$^1$, W. Adam Phelan$^2$, J. F. DiTusa$^2$, \\
D. Alan Tennant$^{4,5}$, Zhiqiang Mao$^{3,6}$\\
\\
\noindent \normalsize{$^{1}$Neutron Scattering Division, Oak Ridge National Laboratory, Oak Ridge, TN 37831, USA}\\
\noindent \normalsize{$^{2}$ Department of Physics and Astronomy, Louisiana State University, Baton Rouge, LA 70803,USA}\\ 
\normalsize{$^{3}$ Department of Physics and Engineering Physics, Tulane University, New Orleans, LA 70118, USA}\\
\noindent  \normalsize{$^{4}$Materials Science and Technology Division, Oak Ridge National Laboratory,Oak Ridge, TN 37831, USA}\\
\noindent \normalsize{$^{5}$ Shull Wollan Center, Oak Ridge National Laboratory, Oak Ridge, TN 37831, USA}\\
\normalsize{$^{6}$ Department of Physics, Pennsylvania State University, University Park, PA 16802, USA}\\
\\
\normalsize{$^{+}$Equal contribution.}\\
\normalsize{$^\ast$To whom correspondence should be addressed; E-mail:zhangq6@ornl.gov. }
}

\date{\today}

\begin{document} 


\baselineskip24pt


\maketitle


\begin{sciabstract}
 Magnetic topological semimetals have attracted intense attention recently since these materials carry a great promise for potential applications in novel spintronic devices. Here, we report an intimate interplay between lattice, Eu magnetic order and topological semimetallic behavior in Eu$_{1-x}$Sr$_{x}$MnSb$_{2}$ driven by nonmagnetic Sr doping on magnetic Eu site. Different types of Eu spin reorientations are controllable by the Sr concentration, temperature or magnetic field, and coupled to the quantum transport properties of Dirac fermions generated by the 2D Sb layers. Our study opens a new pathway to achieving exotic magnetic order and topological semimetallic state via controlling spin reorientation. The effective strategy of substituting rare-earth site by nonmagnetic element demonstrated here may be applicable to the AMnCh$_{2}$ (A=rare-earth elements; Ch=Bi/Sb) family and a wide variation of other layered compounds involving spatially separated rare-earth and transition metal layers. 
 
\end{sciabstract}


\section*{Introduction}

Dirac/Weyl semimetals have attracted intense research interest due to their exotic quantum phenomena as well as promising applications in more energy efficient next-generation electronic devices.\cite{Xu613,Lu622,Lv2015} Compared with many non-magnetic semimetals, magnetic Dirac/Weyl semimetals are especially attractive since the coupling of Dirac/Weyl fermions to additional spin degree of freedom may open up a new avenue to tune and control resulting quantum transport properties.\cite{Armitage2018,Wan2011,LiuEn2018,Liu2017}
Therefore, understanding the coupling between magnetism and topological semimetallic behavior has emerged as a forefront topic in studying magnetic semimetals for both fundamental research and promising applications. The determination of the structural, magnetic and electronic phase diagram in magnetic semimetals is a key to unveiling not only the origin of the topological semimetallic behavior but also the underlying physics associated with the coupling among lattice, magnetism and semimetallic properties. Such phase diagram will also form the basis from which a microscopic theory of such couplings can be established. Furthermore, it is of great interest to explore new strategies to tune quantum transport properties by a spin order in a controllable way. 

A large family of ternary AMnCh$_{2}$ ``112'' compounds (A =alkali earth/rare earth elements, C = Bi or Sb)\cite{Wollesen1996,Liu2017,Borisenko2019,Masudae1501117,Joonbum2011} are interesting since a few of them have been reported to be magnetic Dirac semimetals where the Bi or Sb layers may host relativistic fermions. AMnCh$_{2}$ (A=Ce, Pr, Nd, Eu, Sm; C=Bi or Sb)\cite{Masudae1501117,Yi2017,Soh2019} possesses two magnetic sub-lattices, formed by the magnetic moments of rare-earth A and Mn respectively, in contrast with other compounds showing only Mn magnetic lattice in this family. The conducting Bi/Sb layers and the insulating magnetic Mn-Bi(Sb) and Eu layers are spatially separated, which makes them good candidates to explore the possible interplay between likely Dirac fermions and magnetism. For EuMnBi$_{2}$, both Eu and Mn moments point to the out-of-plane direction and generate two AFM lattices in the ground state \cite{Masudae1501117}. Previous studies have also shown that when the Eu AFM order undergoes a spin-flop transition in a moderate field range, the interlayer conduction is strongly suppressed, thus resulting in a stacked quantum Hall effect \cite{Masudae1501117}. Interestingly, EuMnSb$_{2}$ exhibits distinct properties from EuMnBi$_{2}$ and  conflicting results have been reported \cite{Yi2017,Soh2019,Gong2020}. 
The magneto-transport properties reported by Yi et al. \cite{Yi2017} are not indicative of a Dirac semimetallic state. But the magneto-transport properties observed by Soh et al. \cite{Soh2019} and linear-band dispersion near the Fermi level probed in Angle-Resolved Photoemission Spectroscopy (ARPES) measurements suggest that EuMnSb$_{2}$ may be a Dirac semimetal\cite{Soh2019}. However, no nontrivial Berry phase has been reported on EuMnSb$_{2}$ to support this argument. The magnetic structure of EuMnSb$_{2}$ is also thought to be distinct from that of EuMnBi$_{2}$, with controversial reports on Eu and Mn moments being perpendicular \cite{Soh2019} or canted to each other \cite{Gong2020}. It is therefore important to resolve the controversial magnetic and physical properties of EuMnSb$_{2}$ and to explore whether EuMnSb$_{2}$ and its derivatives could host Dirac fermions. Furthermore, given there are two magnetic sublattices of Eu and Mn with an expected 4$f$-3$d$ coupling between them, the chemical substitution of Eu by nonmagnetic element may be a good strategy for exploring the tunability of magnetism and its possible coupling to transport and magneto-transport properties.

Here, we report a rich phase diagram of crystal structure, magnetism and electronic properties of Eu$_{1-x}$Sr$_{x}$MnSb$_{2}$ established through comprehensive studies using single crystal X-ray diffraction, neutron scattering, magnetic and high-field transport measurements. From this phase diagram, we reveal an intricate interplay between structure, magnetism and quantum transport properties of relativistic fermions. We find the chemical substitution of magnetic Eu by nonmagnetic Sr induces dramatic changes in the structure, magnetism and electronic transport properties. The change of Sr concentration, magnetic field and temperature lead to complex magnetic states with various Eu spin-reorientations, which are coupled to the transport and magneto-transport properties. For x$\geq$0.5, the topological semimetallic behavior appears as indicated by the nontrivial Berry phases and originates from the Sr doping on Eu. Furthermore, the Eu spin canting toward the out-of-plane direction driven by Sr doping decreases the intralayer conductivity but significantly increases the interlayer conductivity between Sb layers that host Dirac fermions. We demonstrate that substituting rare-earth site by nonmagnetic element provides a new strategy to achieve exotic magnetic and topological states via controlling spin reorientations by tuning the competition of 4$f$-3$d$ and A-A magnetic couplings, in layered compounds involving spatially separated rare-earth and transition metal layers.

 \section*{Results and discussion}

 Both single crystal x-ray and neutron diffraction reveal that $x=0$ in Eu$_{1-x}$Sr$_{x}$MnSb$_{2}$ crystallizes in a tetragonal structure with space group $P4/nmm$ (Fig. 1 (a)), with an existing Mn difficiency, i.e., EuMn$_{0.95}$Sb$_{2}$. The structural parameters of EuMnSb$_{2}$ obtained from the single crystal x-ray diffraction at 293 K are summarized in Table SI and SII. Note that the structure of EuMn$_{0.95}$Sb$_{2}$ is similar to CaMnBi$_{2}$, but different from the space group $I 4/mmm$ of tetragonal EuMn$_{0.95}$Bi$_{2}$\cite{Masudae1501117} and the orthorhombic structure in previous reports on EuMnSb$_{2}$\cite{Yi2017,Soh2019,Gong2020}. The Sr-doped Eu$_{1-x}$Sr$_{x}$MnSb$_{2}$ (x=0.2, 0.5 and 0.8), however, show a clear lattice distortion and crystallize in the orthorhombic structure with space group $Pnma$ with doubled unit cell along out-of-plane direction (Fig. 1 (b)), similar to SrMnSb$_{2}$ \cite{Liu2017}. The  structural parameters of Eu$_{1-x}$Sr$_{x}$MnSb$_{2}$ (x=0, 0.2, 0.5 and 0.8) at 5 K obtained from the fits to neutron diffraction data  are summarized in Table I. It can be seen that the Sr doping induces a slight decrease of out-of-plane lattice constant and an increase of in-plane lattice constants.  More details of the determination of crystal structures of all the Eu$_{1-x}$Sr$_{x}$MnSb$_{2}$ compounds can be found in Supplemental Information.

 \begin{figure} \centering \includegraphics [width = 0.85\linewidth] {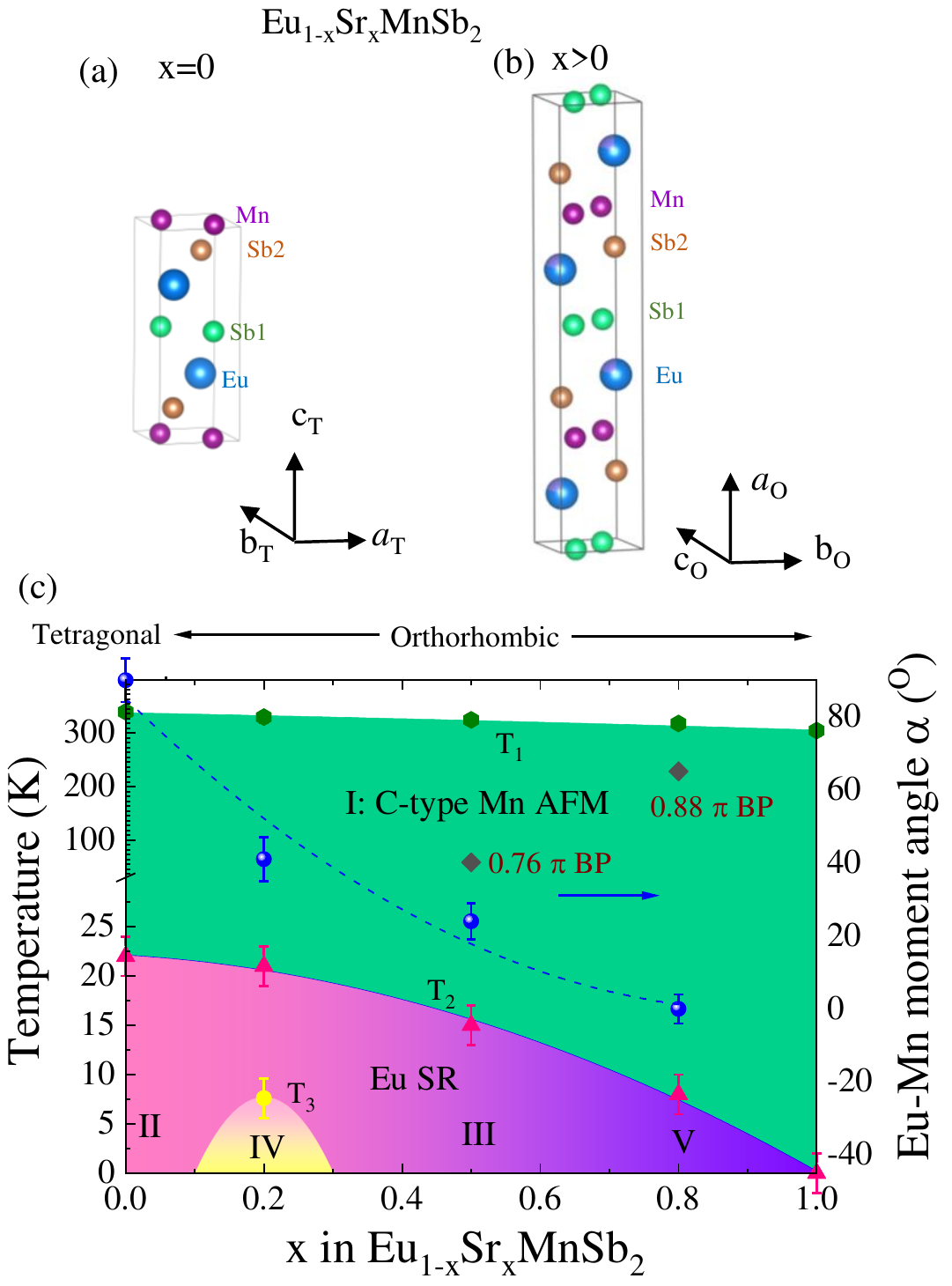}
\caption{(color online) (color online) A view of the crystallographic structure of Eu$_{1-x}$Sr$_{x}$MnSb$_{2}$ for (a) x=0 and (b)x$>$0. (c) Composition phase diagram of Eu$_{1-x}$Sr$_{x}$MnSb$_{2}$ on the structural and magnetic transitions, Eu-Mn moment angle $\alpha$ and nontrivial Berry phase (BH) extracted from the fits to $\rho_{in}$. $T_{1}$, $T_{2}$, and $T_{3}$ label the magnetic transition temperatures. The distinct magnetic structures in I-V regions are displayed in Fig. 3(d-f). The evolution of the  violet color illustrates the gradual decrease of the Eu-Mn moment angles. A higher Eu canting angle of ($90^{o}-\alpha$), i.e., a smaller $\alpha$, is accompanied with the stronger quantum SdH oscillations. The electronic transport properties are also coupled to Eu order at $T_{2}$ and $T_{3}$.  The non-trivial Berry phases indicative of Dirac semimetallic behaviors emerge with $x\geq $0.5.}
\label{fig:structure} 
\end{figure}

 To study the magnetic properties, we measured the susceptibility $\chi$ versus temperature ($T$) and magnetization versus magnetic field $H$ parallel to out-of-plane and in-plane directions, as shown in Fig. 2(a-f). There is no anomaly at high temperatures in $\chi$(T) curves for these four compounds. For x=0, upon cooling below $\approx 20 K$, $\chi_{c}$ increases slightly and $\chi_{ab}$ decreases rapidly, suggesting an AFM order (most likely Eu spins) with moment within the $ab$ plane.  $M_{c}$ is linearly dependent with $H$ at both 3 K and 50 K. Nevertheless, at 3 K, $M_{ab}$ first increases linearly at low fields and starts to increase more rapidly at $\approx 1.7 $ T, indicating a weak modification of the magnetic structure by field. In x=0.2, both $\chi_{a}$ and $\chi_{bc}$ decrease below $\approx 19 $ K, implying that Eu spins may form a canted AFM order. With further decreasing temperature below $\approx 7.5 $ K, $\chi_{a}$ increases and $\chi_{bc}$ decreases anomalously, indicative of another magnetic transition. Both $M_{a}$ and $M_{bc}$ show a linear increase to indicate AFM order at 20 K and 300 K. At 2 K, a weak modification of the magnetic structure occurs at $\approx$ 1.4 and 2.1 T in $M_{bc}$ and $M_{a}$, respectively. In x=0.5, only one anomaly at lower temperature $\approx 17 $ K is observed. As for x=0.8,  $\chi_{bc}$ keeps increasing but $\chi_{a}$ decreases rapidly upon cooling below 8 K, showing the opposite behavior to x=0. This implies that the Eu moment may mainly point to out-of-plane $a$ direction in x=0.8. While both $M_{bc}$ and $M_{a}$ show a typical AFM behavior, a metamagnetic transition occurs with H//a at 2 K. Nevertheless, with H//bc, magnetization exhibits a rapid increase at low fields and then increases at a decreased slope at higher fields, signifying a ferromagnetic (FM) signal. A likely scenario is that Sr doping induces the appearance of a minor FM phase in the Mn sublattice separated from major AFM phase in x=0.8, as reported for the end compound SrMnSb$_{2}$ \cite{Zhang2019}.

\begin{table} 
\centering
  \setlength{\abovecaptionskip}{0pt}%
\setlength{\belowcaptionskip}{10pt}%
\caption{Structural parameters of Eu$_{1-x}$Sr$_{x}$MnSb$_{2}$ with x=0, 0.2, 0.5, and 0.8 at 5 K obtained by the fits to the single crystal neutron diffraction data. For x=0, space group:  $P4/nmm$. Atomic positions: Eu(2c): (0.25, 0.25, $z$), Mn(2a): (0.75, 0.25, 0), Sb$_{1}$(2b): (0.75, 0.25, 0.5), Sb$_{2}$(2c):(0.25, 0.25, $z$). For x$>$0 compounds: Space group: $Pnma$. Eu/Sr(4c): ($x$, 0.25, $z$), Mn(4c): ($x$, 0.25, $z$), Sb$_{1}$(4c):($x$, 0.25, $z$), Sb$_{1}$(4c): ($x$, 0.25, $z$).} 

\setlength{\tabcolsep}{10pt} 
\renewcommand{\arraystretch}{0.8}
\begin{tabular}{cc|c|cc|c|c}
\hline\hline
 &  &  x=0 & \multicolumn{2}{c|}{x=0.2}&  x=0.5 &  x=0.8   \\
\hline
\multicolumn{2}{c|}{lattice constants} &      &        &     &    &   \\
 \multicolumn{2}{c|}{$a$}  &   4.343(6)   &  \multicolumn{2}{c|}{22.348(3) }   &  22.27(42) &  22.28(41) \\
       \multicolumn{2}{c|}{$b$}  & 4.343(6)   &  \multicolumn{2}{c|}{4.347(5)}  &  4.411(14)   & 4.412(14)  \\
      \multicolumn{2}{c|}{$c$} &  11.169(13)  &  \multicolumn{2}{c|}{ 4.383(4)}  &  4.434(24)  & 4.438(28)   \\
\hline
atom      &    &     &       &        &      &     \\
Eu       &   $z$ & 0.729(5)   &  $x$     &       0.113(4)&     0.113(5)&  0.112(4)   \\
         &     &          &  $z$     &      0.781(5) &    0.789(7)  & 0.806(3) \\
Mn       &     &          & $x$     &      0.253(7) &      0.249(4)   &  0.242(4)    \\
         &     &          & $z$     &      0.323(3) &   0.279(7) &  0.292(4)\\
         
Sb$_{1}$ &    &     &   $x$   &      0.0019(8)  &  0.0011(7)  &  0.0042(9)     \\
         &    &          &   $z$    &      0.233(6)  & 0.264(4) &  0.298(7)      \\
Sb$_{2}$ & $z$   &   0.156(7)       &   $x$    &      0.324(5) &  0.325 (5) & 0.324(6)      \\
         &    &           &   $z$    &     0.829(5) & 0.768(4)     &0.818(5) \\
\hline
Reliable factors   &      &           &      &        &      \\
       \multicolumn{2}{c|}{ $R_{f}$} &   8.75        &      &  6.67 & 6.18    &  7.59    \\
          \multicolumn{2}{c|}{ $\chi^{2}$}  &   0.28        &      & 0.28 & 1.21     &   0.83   \\
\hline\hline            
\end{tabular}
\label{tab:Refined structure}
\end{table}

 Single crystal neutron diffraction was employed to determine the complicated magnetic structures of Eu$_{1-x}$Sr$_{x}$MnSb$_{2}$ below $\approx$ 340 K (see details in Supplemental Information). 
 The refined moments, Mn-Eu canting angle and reliability factors of the refinements on the neutron data after neutron absorption correction are summarized in Table II. Figure 3 (a-c) show the temperature dependence of a few representative nuclear and/or magnetic reflections of Eu$_{1-x}$Sr$_{x}$MnSb$_{2}$. For parent x=0, an appearance of pure magnetic peak (100)$\rm{_{T}}$ below $T_{1}\approx 330 $ K indicates one magnetic transition. The absence of an anomaly at $T_{1}$ in susceptibility measurements may be ascribed to the possible strong spin fluctuations above $T_{1}$ that tend to smear out any anomaly in the susceptibility as in other Mn-based compounds \cite{Liu2017,Zhang2015,Zhang2016}. For $T<T_{1}$, a C-type AFM order of Mn spins with moment along the $c$ axis is determined without Eu ordering, as illustrated in the left panel of Fig. 3 (d). Upon cooling below $T_{2}\approx 22 $ K, there is an increase of magnetic peak intensities such as (100)$\rm{_{T}}$ and (101)$\rm{_{T}}$ with ${\bf k} =$ (0,0,0)$\rm{_{T}}$ and simultaneously, new magnetic reflections with a propagation vector ${\bf k} =$ (0,0,1/2)$\rm{_{T}}$ from Eu sublattice appear. Interestingly, we observed strong magnetic peaks (0,0,L/2)$\rm{_{T}}$ (L= odd number) (see inset of Fig. 3(a)). This excludes the Eu moments pointing in the out-of-plane axis seen in EuMnBi$_{2}$ \cite{Masudae1501117}.
The determined magnetic structure for $T<T_{2}$ is shown in the right panel of Fig. 3 (d).  Whereas Mn still has a C-type AFM order with moment along the $c$ axis, the ``+ + - -" Eu spin ordering with moment along the $a$ axis breaks magnetic symmetry along the $c$ axis and leads to observed magnetic reflections with ${\bf k} =$ (0,0,1/2)$\rm{_{T}}$. At 4 K, the perpendicular Mn and Eu moments are found to be 4.63(21) and 5.25(43) $\mu_{B}$, respectively, indicative of Mn$^{2+}$ ($S=5/2$) and Eu$^{2+}$ ($S=7/2$). Note that the magnetic structure determined here is different from the ''+-+-" A-type Eu order proposed on the basis of diffraction experiments on polycrystalline powder sample EuMnSb$_{2}$ since no ${\bf k} =$ (0,0,1/2)$\rm{_{T}}$ magnetic peaks were observed below $T_{2}$. No Eu canting proposed in the \textit{Ref.} 12 is found for $T<T_{2}$ in our crystal, the details of which can be found in Supplemental Information.

   \begin{figure} \centering \includegraphics [width = 1\linewidth] {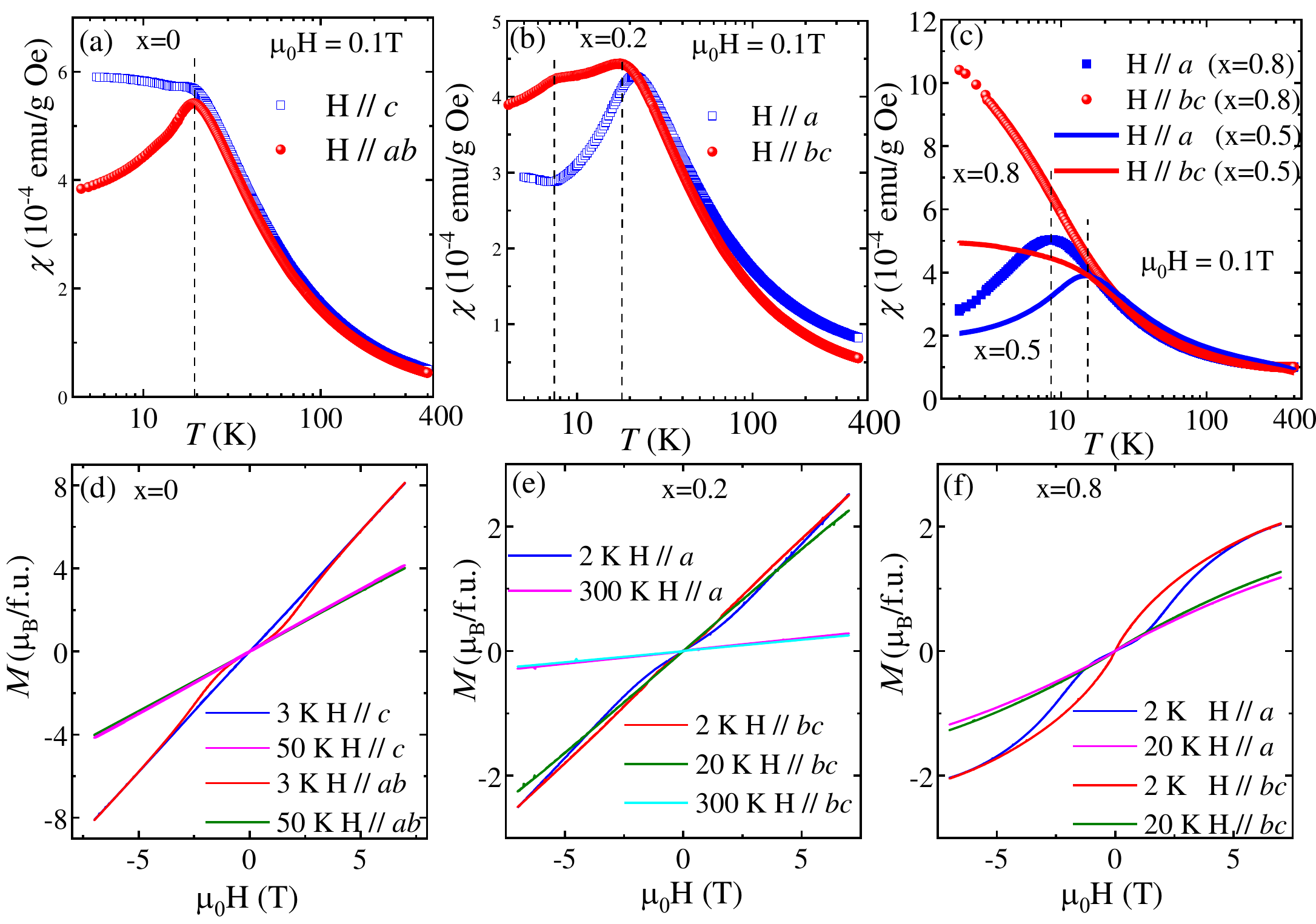}
\caption{(color online) (color online) (color online) Temperature dependence of susceptibility for (a) x=0 with magnetic field of 0.1 T parallel to out-of-plane $c$ and in-plane $ab$ directions, (b) x= 0.2, and (c) x=0.5 and 0.8, with field parallel to out-of-plane $a$ and in-plane  $bc$ directions. The vertical lines indicate the locations of the magnetic transition temperatures. Field dependence of magnetization at different temperatures for (d) x=0, (e) x=0.2 and (f) x=0.8.  }
\label{fig:structure} 
\end{figure}

In x=0.2, we observed pure magnetic peaks (010)$\rm{_{O}}$ and (100)$\rm{_{O}}$, corresponding to (100)$\rm{_{T}}$ in tetragonal notation, below $T_{1}\approx$ 330 K and determined a similar C-type AFM order with {\bf k} $=$ (0,0,0)$\rm{_{O}}$ (see left panel in Fig. 3(e)). Upon cooling below $T_{2}\approx$ 21 K, new magnetic peaks indexed by (H,K,L) (H=odd integers) for instance (700)$\rm{_{O}}$, corresponding to (0 0 3.5)$\rm{_{T}}$, are observed (see inset of Fig. 3 (b)). All the magnetic peaks can be described by ${\bf k} =$ (0,0,0)$\rm{_{O}}$ in orthorhombic notation due to doubled unit cell as compared to x=0.
In $T_{2}<T<T_{1}$, we found a canted and noncollinear Eu spin order confined within $ac$ plane with ``+ + - -' component along $c$ axis and ``+ - + -'' component along $a$ axis, coexisting with the C-type Mn AFM order with moment along the $a$ axis (see the middle panel in Fig. 3 (e)). Note that such canted Eu order model is not applicable in the corresponding $T<T_{2}$ in parent x=0. At 10 K, the canting angle between Mn and Eu is 41(9) $\rm^{o}$. 
Interestingly, with further decreasing temperature below $T_{3}\approx$ 7 K, there is a decrease of the (300)$\rm{_{O}}$ peak intensity, with a concurrent increase of intensity of the nuclear peak (600)$\rm{_{O}}$. This strongly indicates a Eu spin-reorientation transition to a Eu spin order without a magnetic symmetry breaking along the $a$ axis. While C-type Mn order is unchanged, a canted and collinear magnetic structure with A-type ``+ - + -'' Eu spin order along both $a$ and $c$ axes is determined with a Mn-Eu canting angle of 40(7)$\rm^{o}$ at 4 K, as shown in the right panel of Fig. 3(e).

  \begin{table}
\centering
 \fontsize{9}{12}\selectfont

\setlength{\abovecaptionskip}{0pt}%
\setlength{\belowcaptionskip}{10pt}%
\caption{Refined magnetic moments, Mn-Eu angles and reliable factors of Eu$_{1-x}$Sr$_{x}$MnSb$_{2}$ with x=0, 0.2, 0.5, and 0.8 at different temperatures.}
\setlength{\tabcolsep}{10pt}
\renewcommand{\arraystretch}{0.7}
\tabcolsep=0.1cm

\begin{tabular}{cc|cc|ccc|cc|cc|}
 \hline\hline
            &              &\multicolumn{2}{c|}{x=0} &\multicolumn{3}{c|}{x=0.2} & \multicolumn{2}{c}{x=0.5}& \multicolumn{2}{c}{ x=0.8}     \\ \cline{2-11}  
           &  $T$~(K)      &    170 &5              &     60   & 10       & 4      &    50 &  5                 &        \multicolumn{2}{c}{  5}     \\    
           \hline
{Mn moments}& $M_{c}(x=0)$ & 2.99(29)& 4.63(21)     &           &           &       &        &                   &  \multicolumn{2}{c}{ }                        \\
           &  $M_{a}(x>0)$ &         &               & 3.70(46) &3.66(32) &3.75(45) & 3.74(15) & 3.76(17)        &  \multicolumn{2}{c}{3.80(22)}                \\
  \hline
{Eu moments} & $M_{a}$   &          &               &          & 4.08(34) & 3.89(69) &         & 4.84(55)        &  \multicolumn{2}{c}{5.17(62)}                  \\
           & $M_{b}$   &            &               &          &          &         &          &                 &   \multicolumn{2}{c}{ }                        \\
          & $M_{c}$   &               &5.25(43)    &            & 3.52(34) &3.30(86)  &         & 2.23(29)        &                        \\
           & $|M_{total}|$ &   & 5.25(43)          &           & 5.38(34)  & 5.26(50)  &       & 5.32(50)         & \multicolumn{2}{c}{5.17(62)}               \\
          \hline
{Mn-Eu moment angle($\rm^{o}$)} & & &90        &            &  41(9)   & 40(7)      &       & 24(8)            & \multicolumn{2}{c}{0}                   \\
    \hline
{Reliable factors}&$R_{F}({\bf k}=(0,0,0))$ & 9.53  & 8.75 &7.64 &7.55 &   6.67   &  5.32   &  6.18             & \multicolumn{2}{c}{7.59}                    \\
           & $\chi_{2}({\bf k}=(0,0,0))$    & 0.27   & 0.28 & 0.13&  0.29& 0.28    & 0.31    & 1.26              & \multicolumn{2}{c}{0.83}                   \\
           & $R_{F}({\bf k}=(0,0,1/2)_{T})$      &         & 8.93 &   &       &        &         &                   &  \multicolumn{2}{c}{ }                        \\
           & $\chi_{2}({\bf k}=(0,0,1/2)_{T})$   &          & 0.26 &   &     &         &         &                   &  \multicolumn{2}{c}{ }                       \\
  \hline \hline            
\end{tabular}
\label{tab:Refined_moments}
\end{table}

In x=0.5, both the (010)$\rm{_{O}}$ and (001)$\rm{_{O}}$ magnetic peaks appear below $T_{1}$. Upon cooling below $T_{2}\approx$ 15 K, the (010)$\rm{_{O}}$ peak intensity further increases while there is no obvious change in the (001)$\rm{_{O}}$ (see Fig. 3(c) and Fig. S5(a-b)). Furthermore, there is an increase of magnetic peak intensity (300)$\rm{_{O}}$ but no obvious change in peak intensities of (200)$\rm{_{O}}$ or (600)$\rm{_{O}}$. These features are similar to those in x=0.2. We determined the similar magnetic structures in x=0.5, as shown in the left and middle panels in Fig. 3 (e) for $T_{2}<T<T_{1}$ and $T_{3}<T<T_{2}$, respectively. Note that the canting angle between Eu and Mn moments decreases to $\approx$ 24$\rm^{o}$ at 5 K. As for x=0.8, a Mn magnetic transition occurs at a temperature $T_{1}\approx$ 330 K as identified from the intensity of (010)$\rm{_{O}}$ and a C-type Mn order is determined (see the left panel of Fig. 3(f)). Another increase of (010)$\rm{_{O}}$ is found below  $T_{2}\approx$ 7 K. There is no appearance of magnetic scattering at the (300)$\rm{_{O}}$ and (200)$\rm{_{O}}$ or (600)$\rm{_{O}}$  Bragg positions below  $T_{2}$ (see Fig. 3 (c) and Fig. S5(c-d) in SI), indicating that Eu moments may point to the $a$ axis. We found a coexistence of Mn C-type AFM order with the ''+ - + -" Eu order with reoriented moments along the same the $a$ axis as the Mn moment (see right panel of Fig. 3 (f)), consistent with susceptibility measurements.

Next, we present the evolution of electronic transport properties with the Sr doping in 
Eu$_{1-x}$Sr$_{x}$MnSb$_{2}$. As shown in Fig. 4 (a-c), both in-plane ($\rho_{in}$) and out-of-plane resistivity ($\rho_{out}$) exhibit metallic transport properties. At 2 K, the $\rho_{out}/\rho_{in}$ reaches 128, 198 and 322, for x=0, x=0.2 and x=0.8, respectively. The rapid increase of this a ratio indicates that Sr doping reinforces the quasi-2D electronic structure. In x=0, the slope of resistivity in $\rho_{out}$ and $\rho_{in}$ decreases below $T_{2}$, indicative of a correlation between the emergence of Eu order and transport properties, which reveals that the in-plane Eu ``+ + - - " order increases the resistivity for both $\rho_{out}$ and $\rho_{in}$. However, in x=0.2 below $T_{2}$, a rapid decrease in $\rho_{out}$ and increase in $\rho_{in}$ suggest that the Eu canting to the $a$ axis with ``+ - + - " component increases significantly the interlayer conductivity along the $a$ direction between Sb layers but suppresses the intralayer conductivity in the $bc$ plane, showing a different effect on the transport properties as compared to the sole in-plane Eu order in x=0. Below $T_{3}$, there is no obvious change in the out-of-plane resistivity, but an anomalous decrease of the in-plane resistivity is observed. This can be well interpreted from the SR of Eu from noncollinear to collinear order. The out-of-plane Eu order is kept to be ``+ - + - ", which is expected not to influence the interlayer conductivity. In contrast, the reorientation of the in-plane component from ``+ + - - " to ``+ - + -'' induces the anomalous increase in the intralayer conductivity. As for x=0.8, the Eu ordering does not influence the resistivity obviously below $T_{2}$, which is understood due to a small portion of Eu occupancy $\approx$ 20 $\%$. Thus, our results reveal an intimate coupling between Eu magnetic order and transport properties in Eu$_{1-x}$Sr$_{x}$MnSb$_{2}$.

 \begin{figure} \centering \includegraphics [width = 1\linewidth] {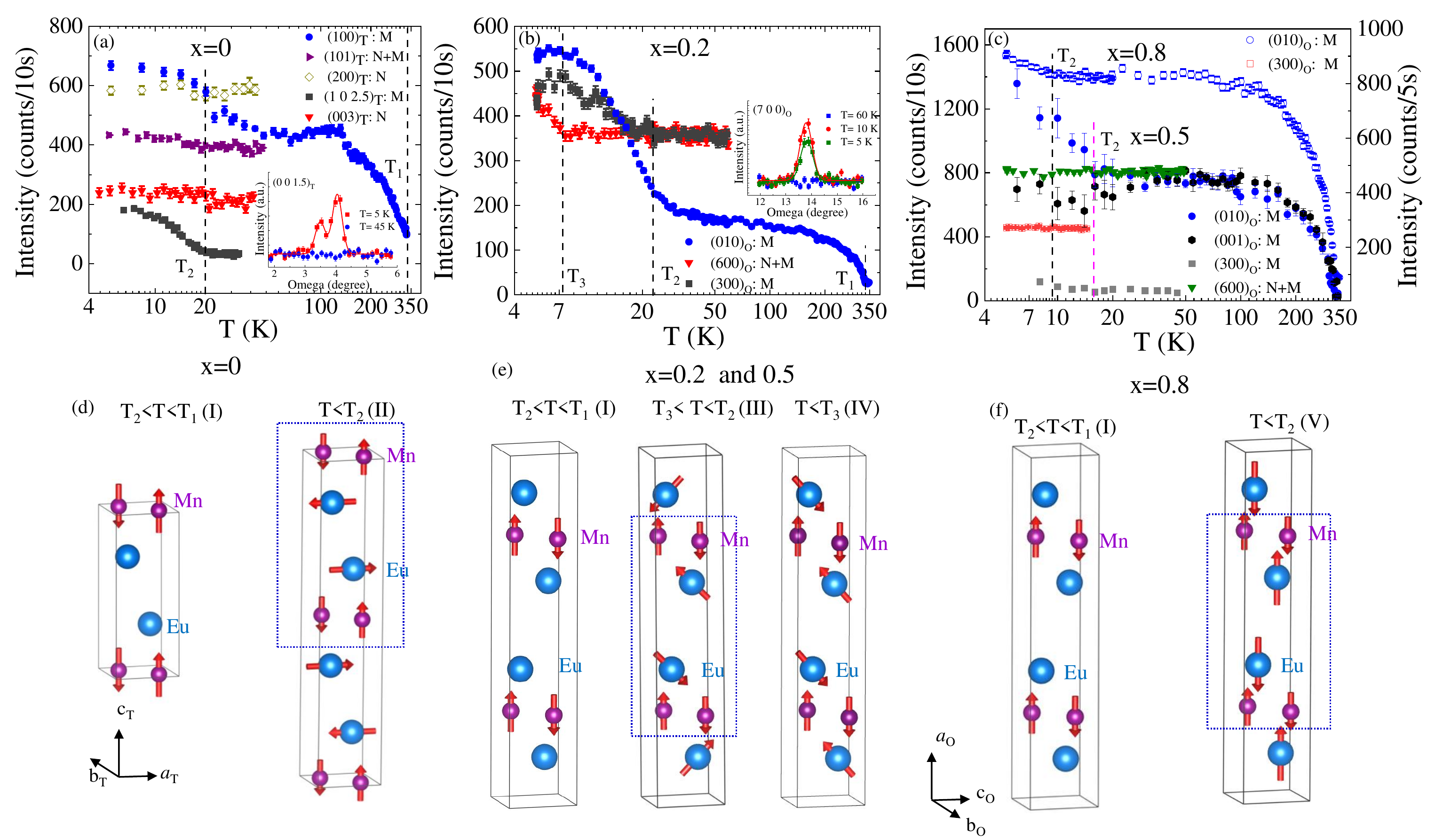}
\caption{(color online) Temperature dependence of intensities at the representative nuclear and/or magnetic peak positions for (a) x=0, (b) x=0.2 and (c) x=0.5 and 0.8. The 2nd weak peak with smaller omaga in the rocking curves for x=0 is due to the presence of another tetragonal domain in the crystal. The data for x=0.5 and x=0.8 compounds are shown by open and solid symbols, respectively. Magnetic structures determined from the fits to the neutron data for (d). x=0, (e) x=0.2 ( all the panels) and 0.5 (only left and middle panels), and (f) x=0.8. The dashed rectangular shows the Mn-Eu-Eu-Mn block where the SR of Eu can be seen.   
}
\label{fig:neutron} 
\end{figure}

Figure 4 (d-f) shows both in-plane and out-of-plane magneto-transport properties ( $MR=[\rho(B)-\rho(0)]/\rho(0)$) under high fields applied along out-of-plane direction. For x=0, the  $\Delta \rho_{out}/\rho_{out}$ is positive, whereas the in-plane $\Delta \rho_{in}/\rho_{in}$ is negative. The magnitudes for both $\Delta \rho_{out}/\rho_{out}$ and $\Delta \rho_{in}/\rho_{in}$ are small, and no strong Shubnikov-de Haas (SdH) oscillations are observed. In x=0.2, weak SdH oscillations are observed in both $\Delta \rho_{out}/\rho_{out}$  and $\Delta \rho_{in}/\rho_{in}$. As field increases, there is a sign switch in the $\Delta \rho_{in}/\rho_{in}$, whereas the $\Delta \rho_{out}/\rho_{out}$ remains positive. Remarkably, at 1.8 K that is below $T_{3}$, a large jump in $\Delta \rho_{out}/\rho_{out}\approx$ 3000 $\%$ occurs for $B\approx 18$ T. Such enhanced $\Delta \rho_{out}/\rho_{out}$ is ascribed to a field-induced magnetic structural transition. Since this phenomena does not occur in $T>T_{2}$ for example 50 K, the field-induced magnetic transition should not originate from the Mn magnetic sublattice, but be related to the Eu magnetic sublattice, indicative of a vital role that the Eu magnetic order plays in the magneto-transport properties. A very likely origin of the enhanced $\Delta \rho_{out}/\rho_{out}$ is the field-induced Eu SR transition from canted moment direction in $ac$ plane to the $c$ axis while A-type Eu order remains as illustrated in the inset of Fig. 4(e), yielding a strong suppression of interlayer conductivity. Note that this is different from the filed-induced spin flop of ``+ + - -" Eu order from out-of-plane $c$ axis to in-plane direction in EuMnBi$_{2}$ \cite{Masudae1501117}. The increase of the Sr doping level enhances SdH oscillations significantly in both $\Delta \rho_{out}/\rho_{out}$ and $\Delta \rho_{in}/\rho_{in}$ for x=0.5 and 0.8, with a much higher magnitude at high magnetic fields. $\Delta \rho_{out}/\rho_{out}$ reaches remarkable $\approx 18000\%$ at 31.5 T for x=0.8. We further examine the Berry phase (BP) $\phi_{B}$ accumulated along cyclotron orbits and are able to extract $\phi_{B}$ for x=0.5 and 0.8 (see more details in Supplemental information). From the linear fit of the Landau level fan diagram based on the oscillatory resistivity $\rho_{out}$ of x=0.8, we extract the accumulated phase to be 1.14 $\pi$ as shown in the inset of Fig. 4 (f). The accumulated phases obtained from the fits to $\rho_{in}$ yield 0.76 $\pi$ for x=0.5 and 0.88 $\pi$ for x=0.8, as shown in the Fig. S 7(a) and S 7 (b), respectively. All of them are close to a nontrivial $\pi$ Berry phase for a quasi 2D system. The non-trivial Berry phase provides the evidences that x=0.5 and 0.8 harbor relativistic Dirac fermions. Our results clearly show that the substitution of Eu by nonmagnetic Sr induces an appearance of Dirac semimetallic behavior, which is closely associated with the controllable Eu magnetic order. 
       \begin{figure} \centering \includegraphics [width = 1\linewidth] {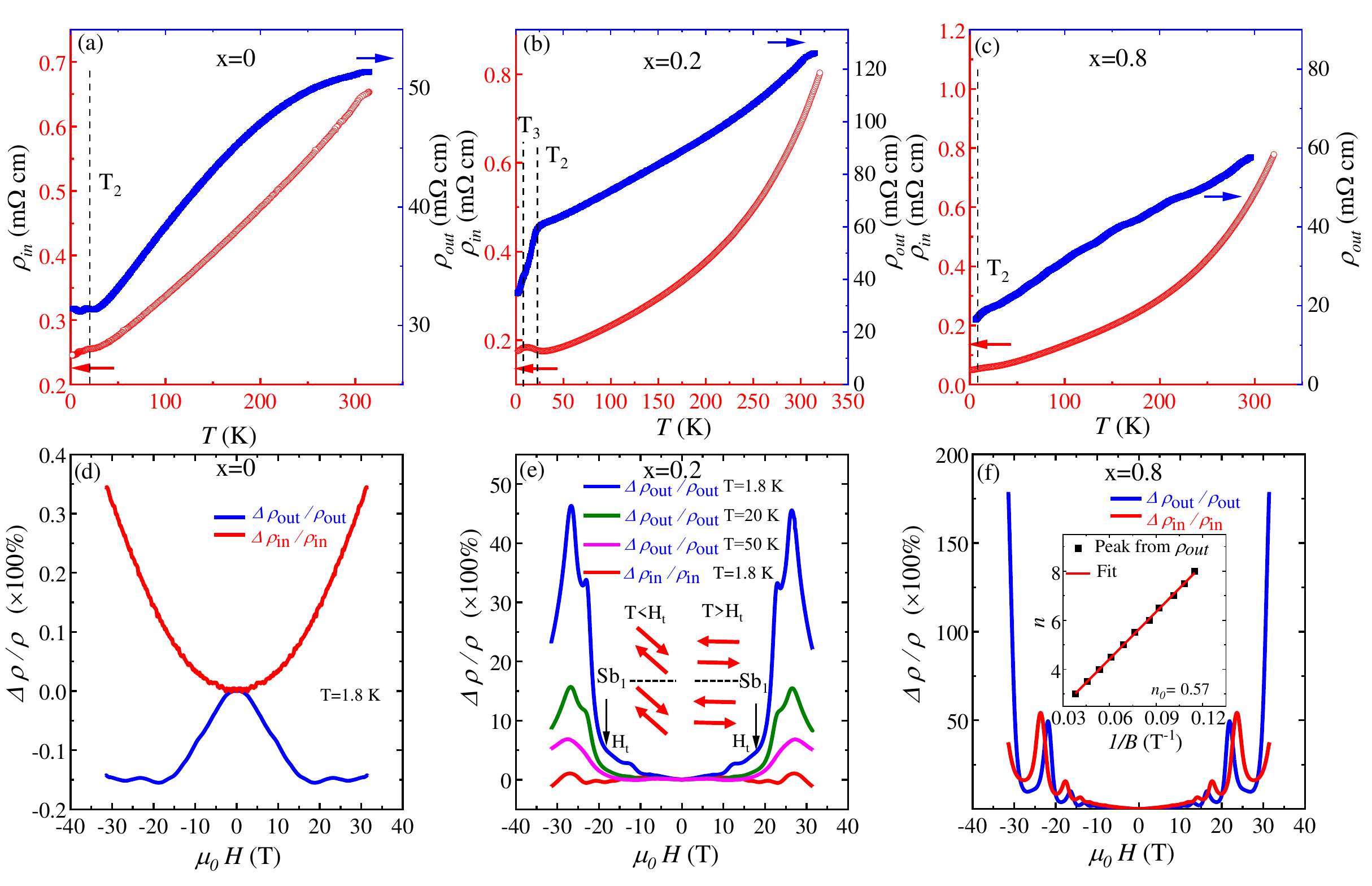}
\caption{(color online) Temperature dependence of resistivity at zero magnetic field for out-of-plane resistivity $\rho_{out}$ and in-plane resistivity $\rho_{in}$ for (a) x=0, (b) x=0.2 and (c) x=0.8. (d-f). Field dependence of out-of-plane magnetoresistance $\Delta \rho_{out}/\rho_{out}$ and in-plane magnetoresistance $\Delta \rho_{in}/\rho_{in}$ for (d) x=0. (e) x=0.2 and (f) x=0.8. The inset of (f) shows the linear fit of the Landau level fan diagram based on the oscillatory resistivity $\rho_{out}$ for x=0.8. The intercept n$_{0}$ is obtained to be 0.44, corresponding to 2$\pi \times$ n$_{0}$=0.88 $\pi$ Berry phase.}
\label{fig:transport} 
\end{figure}

From the combination of single crystal x-ray, neutron, magnetization, and magneto-transport techniques, we are able to establish the structural, magnetic and electronic phase diagram, as shown in Fig. 1 (c). Whereas parent x=0 with Mn difficiency is tetragonal with space group $P4/nmm$, the Sr-doping induces a symmetry breaking to orthorhombic. This is consistent with the previous reports on the orthorhombic structure of Sn doped EuMnSb$_{2}$ \cite{Yi2017} and Mn-overdopped EuMn$_{1.1}$Sb$_{2}$ \cite{Gong2020}. EuMnSb$_{2}$ forms a magnetic structure with perpendicular Mn and Eu moments as ground state and does not exhibit topological semimetallic behavior. Note that the crystal structure, magnetic structure and electronic properties of EuMnSb$_{2}$ are very different from those in EuMnBi$_{2}$\cite{Masudae1501117}. The substitution of Sr on the Eu site in EuMnSb$_{2}$ induces a slight decrease of $T_{1}$ but suppresses $T_{2}$ significantly. Furthermore, the increase of Sr doping drives a gradual Eu SR from the in-plane to the out-of-plane direction and simultaneously, an appearance of the Dirac semimetallic behaviors. A higher Eu canting angle characterized by a smaller Eu-Mn angle is accompanied with the stronger quantum SdH oscillations. Our results show that the Eu spin canting can be driven by the chemical doping, which could interpret the observation of Eu canting in Mn-overdopped EuMn$_{1.1}$Sb$_{2}$ \cite{Gong2020}. It is worthwhile pointing out that Mn overdopping in EuMn$_{1.1}$Sb$_{2}$ does not induce another magnetic transition at $T_{3}$. For the optimized composition x=0.2, the 2nd type of Eu SR from a noncollinear canted spin order to a collinear A-type canted spin order was found at lower temperature. Furthermore, the Eu order at base temperature can be easily tuable by the external magnetic field to result in the 3rd type SR to reorient the canted moment to in-plane $c$ axis. Our phase diagram on  Eu$_{1-x}$Sr$_{x}$MnSb$_{2}$  and comparison to the previous reports \cite{Yi2017,Soh2019,Gong2020} indicate that the structure, magnetic order and electronic properties of EuMnSb$_{2}$ are easily perturbed by the chemical doping on any of the Eu, Mn and Sb sites yielding the compositions deviating from ideal ``112". This is indeed the case for the samples in previous reports\cite{Yi2017,Soh2019,Gong2020}, which may account for the conflicting results on the structure, magnetic and electronic transport properties. Furthermore, our phase diagram clearly shows that the Sr doping on Eu site is the driving force of the Dirac semimetallic behaviors in Eu$_{1-x}$Sr$_{x}$MnSb$_{2}$, which is discussed in terms of the following three aspects here: 1). The Sr doping on Eu site lowers the lattice symmetry and modifies structural parameters as summarized in Table I, which is expected to change the electronic band structure and may help form the Dirac cones; 2). The increase of Sr doping also induces a ferromagnetism as seen in x=0.8 coexisting with the Mn AFM order. Previous DFT calculations on SrMnSb$_{2}$ \cite{Zhang2019} show that the FM order helps locate the Dirac point near the Fermi level in contrast to AFM Mn order which moves the Dirac points away from the Fermi level; 3). The different types of Eu spin reorientations driven by the Sr concentration, temperature or magnetic field influence the electronic transport and magneto-transport properties significantly. Therefore, our phase diagrams demonstrate an intimate interplay between lattice, Mn and Eu magnetism and semimetallic behaviors. The lattice and magnetic information presented in this work are also indispensable for theoretical calculations to explore the microscopic origin of such interplay.  

Finally, we turn to discuss the origins of the complicated magnetic structures, in particular the Sr-doping and temperature induced Eu SR transition in Eu$_{1-x}$Sr$_{x}$MnSb$_{2}$. A common SR of rare earth is that the rare-earth element drives Mn moment parallel to its moment direction once the rare earth spins are ordered with preferred in-plane orientation at low temperatures, as reported in several compounds such as RMnAsO (R=Nd or Ce) \cite{Zhang2015,Emery2011} and RMnSbO (R=Pr or Ce) \cite{Zhang2016,Kimber2010}. However, the Sr doping in Eu$_{1-x}$Sr$_{x}$MnSb$_{2}$ drives a novel Eu SR with moment from in-plane direction to out-of-plane one while Mn moment direction remains along the out-of-plane $a$-axis. The Mn$^{2+}$ moment, which commonly displays very weak single-ion anisotropy as expected for the L=0 of Mn$^{2+}$ ($S=5/2$), favors orientation along the out-of-plane direction \cite{Zhang2015,Zhang2016,Emery2011}, i.e., the $c$ axis in tetragonal structure or the $a$ axis in orthorhombic structure, forming the C-type AFM order in $T_{2} <T<T_{1}$ of Eu$_{1-x}$Sr$_{x}$MnSb$_{2}$. The inplane checker-board-like AFM structure of the C-type order suggests that the NN interaction J$_{1 }$ is dominant, whereas in-plane next-nearest-neighbor (NNN) interaction J$_{2}$ is very weak. In the context of $J_{1}-J_{2}-J_{c}$ model  \cite{Johnston2011}, we conclude that $J_{1}> 0$, $J_{2}<J_{1}/2$
and out-of-plane $J_{c}<0$ with negligible spin frustration in Mn sublattice. With cooling to $T<T_{2}$, Eu-Eu coupling starts to come into play and induces Eu ordering with preferred orientation of Eu$^{2+}$ ($S=7/2$) within in plane \cite{Javier2009,Xiao2009}, either the $ab$ plane in tetragonal structure or the $bc$ plane in orthorhombic structure. Simultaneously, the Eu-Mn coupling also plays an important role by exerting an effective field, which has the tendency to influence the Mn/Eu moment directions. The increase of Sr concentration on Eu site weakens the Eu-Eu coupling and destabilizes the preferred orientation of the Eu spins. Thus, as x increases to 0.2, the effective field by Eu-Mn coupling tends to drive the Eu moment towards to the Mn moment direction. The competition of Eu-Eu and Eu-Mn couplings induces a spin frustration in Eu sublattice and leads to a canted Eu order with moment in $ac$ plane stabilized in $T_{3} <T<T_{2}$. The increase of Sr doping has a tendency to further drive Eu moment tilt towards the $a$ axis due to weakened Eu-Eu coupling, as shown by a decreased Eu-Mn angle for x=0.5. As Sr doping increases to 0.8, the Eu-Mn coupling overwhelms the weak Eu-Eu coupling, which leads to a SR of Eu to the same moment direction as Mn moment. This could account for the unusual Eu SR induced by Sr doping.  As the temperature decreases below $T_{3}$ for x=0.2, a temperature-induced SR transition occurs. This may be ascribed to another type of Eu-Eu coupling that comes into play below $T_{3}$.  This retains ``+ - + -" out-of-plane component but switches in-plane component from "+ + - -'' to ``+ - + -'', leading to collinear A-type AFM order of Eu spins in $T<T_{3}$. Thus, the striking Eu spin reorientation driven by Sr doping and temperature indicates a strong Eu-Mn (4$f$-3$d$) couplings and results from their competitions to Eu-Eu couplings. It is expected that substituting rare-earth A site by nonmagnetic element may be an effective strategy to achieve exotic magnetic and topological states by tuning the competitions between A-Mn and A-A couplings in the the same family of AMnCh$_{2}$ (A=Ce, Pr, Nd, and Sm; Ch=Bi or Sb) compounds\cite{Wollesen1996}, which may be applicable to other layered compounds involving spatially separated rare-ears and transition metal layers.

To summarize, we establish the structural, magnetic and electronic phase diagrams of Eu$_{1-x}$Sr$_{x}$MnSb$_{2}$ and demonstrate a tuning of the quantum transport via controlling various spin reorientations. While EuMnSb$_{2}$ crystallizing in a tetragonal structure is not a semimetal, the doping of nonmagnetic Sr to Eu site induces a lattice symmetry breaking from tetragonal to orthorhombic structure, an unusual Eu spin reorientation to the out-of-plane direction, as well as the appearance of semimetallic behavior. We also found a higher Eu canting angle is accompanied with enhanced quantum transport properties as manifested by strong quantum oscillations. In an optimized composition Eu$_{0.8}$Sr$_{0.2}$MnSb$_{2}$, the Eu order is found to be easily tunable by either temperature or external magnetic field, which induces different types of Eu spin reorientation. The striking Eu spin reorientations driven by Sr doping and temperature are ascribed to the competition between Eu-Mn (4$f$-3$d$) and Eu-Eu couplings that are tuned by the Sr concentration. The variation of Eu spin reorientations strongly affects both intralayer and interlayer conductivity, indicating the strong coupling between magnetism and electronic transport properties. The significant tunability of magnetism due to various Eu spin reorientations in Eu$_{1-x}$Sr$_{x}$MnSb$_{2}$ provides a rare opportunity to discover novel topological states through the magnetism tuning by temperature, magnetic field and chemical modification. 

 \section*{Materials and Methods} 
 
\subsection{Crystal growth}The Eu$_{1-x}$Sr$_{x}$MnSb$_{2}$ (x=0, 0.2, 0.5, 0.8) single crystals were grown using a flux method.  The starting materials with a stoichiometric mixture of Eu/Sr, Mn and Sb elements were put into a small alumina crucible and sealed in a quartz tube in an argon gas atmosphere. The tube was heated to 1050 $\rm^{o}$C for 2 days, followed by a subsequent cooling to 650  $\rm^{o}$C at a rate of 2  $\rm^{o}$C/h. Plate-like single crystals were obtained.  
 
\subsection{Single-crystal x-ray and neutron diffraction measurements and neutron data analysis} A crystal of x=0 was mounted onto a glass fibers using epoxy, which was in turn then mounted onto the goniometer of a Nonius KappaCCD diffractometer equipped with Mo K$\alpha$ radiation ($\lambda$ = 0.71073 $\AA{}$).  After the data collection and subsequent data reduction, SIR97 was employed to give a starting model, SHELXL97 was used to refine the structural model, and the data were corrected using extinction coefficients and weighting schemes during the final stages of refinement.\cite{Altomare1999,Sheldrick2008} To investigate the crystal and magnetic structures, neutron diffraction measurements were conducted at the four circle  neutron diffractometer (FCD), located at the High Flux Isotope Reactor, Oak Ridge National Laboratory. To further distinguish between tetragonal and orthorhombic structures for x=0, neutrons with a monochromatic wavelength of 1.003 \AA{} without $\lambda$/2 contamination are used via the silicon monochromator from (bent Si-331) \cite{Chakoumakos2011}. For other Eu$_{1-x}$Sr$_{x}$MnSb$_{2}$ (x= 0.2, 0.5, 0.8) crystals, we employed the neutrons with a wavelength of 1.542 \AA{} involving ~1.4 $\%$ $\lambda/$2 contamination from the Si-220 monochromator using its high resolution mode (bending 150) \cite{Chakoumakos2011}. The crystal and magnetic structures were investigated in different temperature windows. The order parameter of a few important nuclear and magnetic peaks was measured. Data were recorded over a temperature range of $4 <T<340$ K using a closed-cycle refrigerator available at the FCD. Due to the involvement of the high absorbing Europium in the Eu$_{1-x}$Sr$_{x}$MnSb$_{2}$ crystals, a proper neutron absorption corrections on the integrated intensities of the nuclear/magnetic peaks is indispensable. The dimensions of the faces for each crystal were measured and a Face Index Absorption Correction on integrated intensities was conducted carefully using WinGX package \cite{Louis2012}. The SARAH representational analysis program \cite{WILLS2000} and Bilbao crystallographic server \cite{Perez2015} were used to derive the symmetry-allowed magnetic structures and magnetic space groups. The full data sets at different temperatures were analyzed using the refinement program FullProf suite \cite{Juan1993} to obtain the structure and magnetic structures.

\subsection{Magnetization, and magneto-transport measurements} The temperature and field dependence of the magnetization were measured in a superconducting quantum interference device magnetometer (Quantum Design) in magnetic fields up to 7 T. The transport measurements at zero magnetic field were performed with a four-probe method using Physical Property Measurement Systems (PPMS). The high-field magneto-transport properties were measured in the 31 T resistivity magnets at the National High Magnetic Field Laboratory (NHMFL) in Tallahassee. The magnetic fields were applied parallel to out-of-plane direction to study the in-plane and out-of-plane magnetoresistance. The $\rho_{in}$ samples were made into hall bar shape and the $\rho_{zz}$ samples were in Corbino disk geometry. Berry phase was extracted from the Landau fan diagram. The integer Landau levels are assigned to the magnetic field positions of resistivity minima in SdH oscillations, which correspond to the minimal density of state. 
 
%

%

\section*{Acknowledgments}
Q.Z. J.L, A. P, J. F.D. and Z.M. acknowledge support for the materials preparation and measurements of magnetization, transport and magneto-transport properties and neutron scatting from the US DOE under EPSCoR Grant No. DESC0012432 with additional support from the Louisiana board of regent. D.A.T. is sponsored by the DOE Office of Science, Laboratory Directed Research and Development program (LDRD) of Oak Ridge National Laboratory, managed by UT-Battelle, LLC for the U.S. Department of Energy. (Project ID 9566).  Z.Q.M. acknowledges the support by the US National Science Foundation under grant DMR 1917579. Use of the high flux isotope reactor at Oak Ridge National Laboratory was supported by the U.S. Department of Energy, Office of Basic Energy Sciences, Scientific User Facilities Division. \\
The authors declare that they have no competing financial interests.\\
This manuscript has been authored by UT-Battelle, LLC, under contract DE-AC05-00OR22725 with the US Department of Energy (DOE). The US government retains and the publisher, by accepting the article for publication, acknowledges that the US government retains a nonexclusive, paid-up, irrevocable, worldwide license to publish or reproduce the published form of this manuscript, or allow others to do so, for US government purposes. DOE will provide public access to these results of federally sponsored research in accordance with the DOE Public Access Plan (http://energy.gov/downloads/doe-public-access-plan).

\clearpage

\clearpage

\section*{Supplementary materials}
\setcounter{figure}{0}
\setcounter{table}{0}
\setcounter{page}{0}

\renewcommand{\thepage}{S\arabic{page}} 
\renewcommand{\thesection}{S\arabic{section}}  
\renewcommand{\thetable}{S\arabic{table}}  
\renewcommand{\thefigure}{S\arabic{figure}}

\textbf{I. Single-crystal neutron diffraction study on the  crystal structure of Eu$_{1-x}$Sr$_{x}$MnSb$_{2}$}\\

 \begin{figure} \centering \includegraphics [width = 1\linewidth] {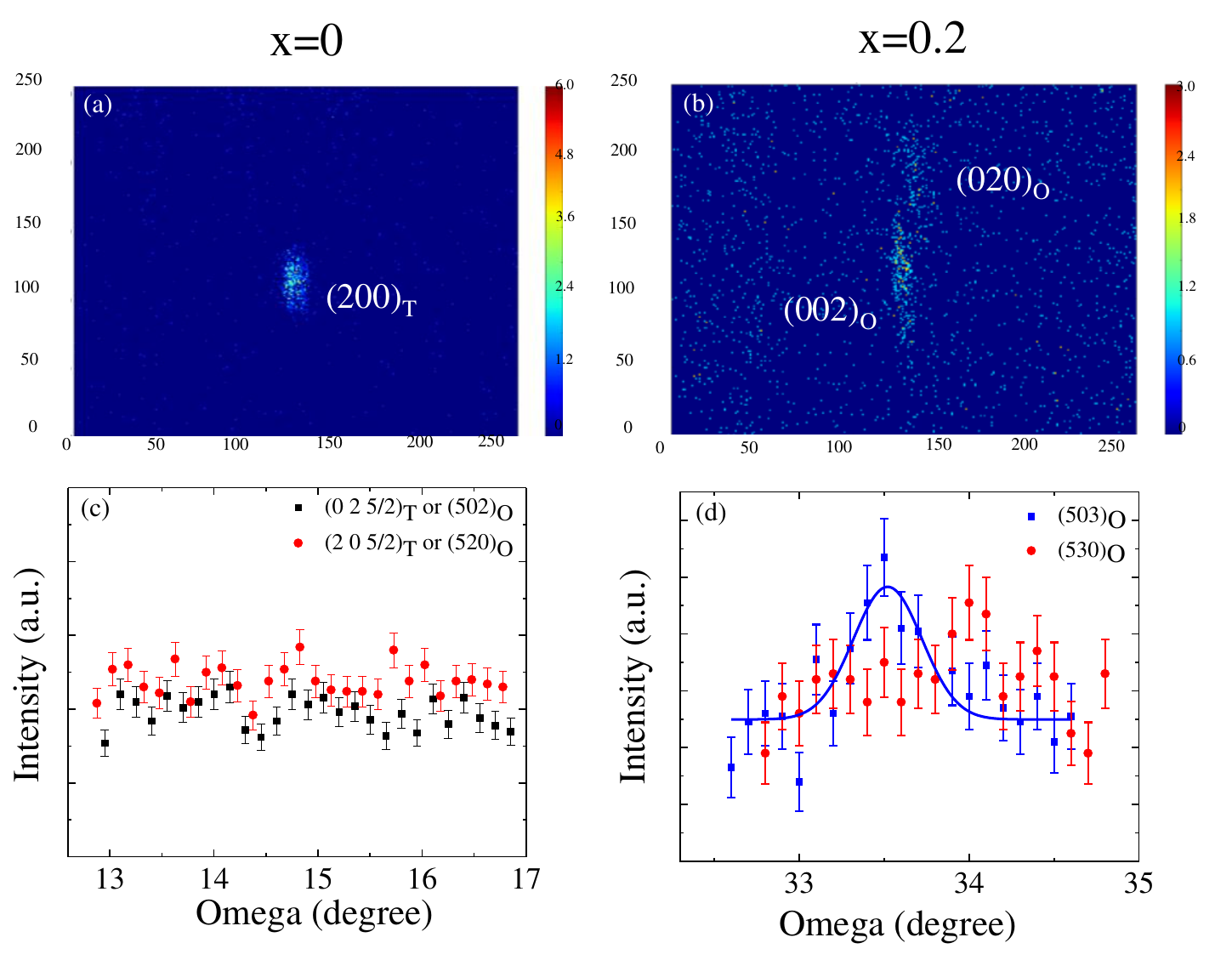}
\caption{(color online)  (a) Observation of single (200)$\rm{_{T}}$ Bragg peak in x=0.2. (b).
Splitting of single (200)$\rm{_{T}}$ peak to (020)$\rm{_{O}}$ and (002)$\rm{_{O}}$ peaks in x=0.2. (c) Absence of (0 2 5/2)$\rm{_{T}}$
and (2 0 5/2)$\rm{_{T}}$ peaks in the rocking curves of x=0. (d). Rocking curves through (503)$\rm{_{O}}$, i.e.,
(0 3 5/2)$\rm{_{T}}$ and (530)$\rm{_{O}}$, i.e., (3 0 5/2)$\rm{_{T}}$) positions in x=0.2    }
\label{fig:structure} 
\end{figure}

Single crystal neutron diffraction was employed to investigate the crystal and magnetic structures in Eu$_{1-x}$Sr$_{x}$MnSb$_{2}$. As shown in Fig. S1 (a), there is one peak for (200)$\rm{_{T}}$ in EuMnSb$_{2}$. However, it shows a clear splitting and becomes two peaks indexed as (020)$\rm{_{O}}$ and (002)$\rm{_{O}}$ in orthorhombic notation indicative of a lattice distortion and intrinsic orthorhombic twinning in x=0.2 (see Fig. S1 (b)). Many Bragg peaks that are allowed by $Pnma$ but forbidden by higher symmetry of $P4/nmm$ were not observed in EuMnSb$_{2}$ from both single crystal x-ray and neutron diffraction measurements. For instance, the Bragg peaks (50L)$\rm{_{O}}$ (L=integer) are allowed but the Bragg peaks (5L0)$\rm{_{O}}$ (L=integer) are forbidden in $Pnma$. Both of them, corresponding to (0 L 5/2)$\rm{_{T}}$ and (L 0 5/2)$\rm{_{T}}$ are forbidden in $P4/nmm$. As displayed in Fig. S1 (c), both (0 2 5/2)$\rm{_{T}}$ and (2 0 5/2)$\rm{_{T}}$ are not observed in EuMnSb$_{2}$. In contrast, x=0.2 exhibits a clear (503)$\rm{_{O}}$ peak with the absence of (530)$\rm{_{O}}$ peak in the rocking curves (see Fig. S1 (d)). Note that a very weak increase of the intensity away from the center of rocking curve of (530)$\rm{_{O}}$ peak arises from the (503)$\rm{_{O}}$ peak in another twinned domain.

\textbf{II. Detailed determination of complicated magnetic structures of Eu$_{1-x}$Sr$_{x}$MnSb$_{2}$\\}

\textbf{1). x=0 }In $T_{2}<T<T_{1}$ of EuMnSb$_{2}$, all of the magnetic reflections can be indexed with the $P4/nmm$ unit cell with a magnetic propagation vector ${\bf k}_{1}$ = (0,0,0)$\rm{_{T}}$. Whereas the (100)$\rm{_{T}}$ Bragg peak is purely magnetic, other peaks for instance (101)$\rm{_{T}}$ and (102)$\rm{_{T}}$ have nuclear and magnetic contributions, consistent with a typical C-type Mn AFM order. In $T<T_{2}$, besides the existence of ${\bf k}_{1} = $ (0,0,0)$\rm{_{T}}$ magnetic peaks, many pure magnetic peaks with a magnetic propagation vector ${\bf k}_{2}$ of (0,0,1/2)$\rm{_{T}}$ appear, indicating the ordering of Eu spins. The SARAH representational analysis program  was employed to determine the symmetry of the allowed magnetic structures for both propagation vectors, as summarized in Table \ref{basis_vector_table_1} and \ref{basis_vector_table_2}. 

 \begin{figure} \centering \includegraphics [width = 1\linewidth] {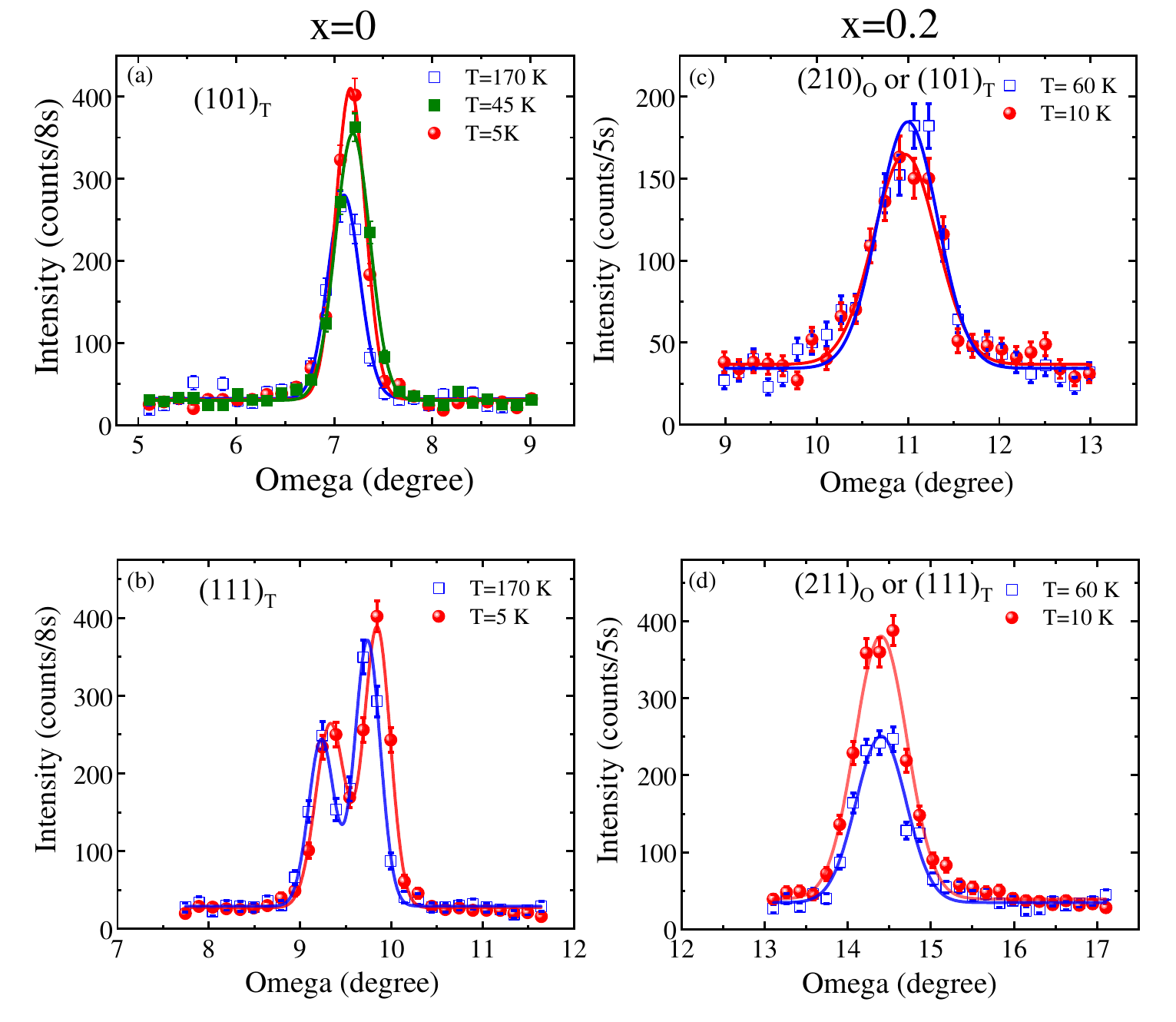}
\caption{(color online)  Rocking curves of (a) (101)$\rm{_{T}}$ and (c) (111)$\rm{_{T}}$ peaks in x=0 at different
temperatures. Rocking curves of the (b) (210)$\rm{_{O}}$, corresponding to (101)$\rm{_{T}}$ and (d) (211)$\rm{_{O}}$,
corresponding to (111)$\rm{_{T}}$ at different temperatures for x=0.2.  }
\label{fig:structure} 
\end{figure}

At 170 K, the integrated intensities of all the nuclear and magnetic reflections can be best fitted employing the $\Gamma_{6}$ magnetic structure with the fits yielding $M_{c}$=2.99(29) $\mu_{B}$.  Thus,  in $T_{2} <T<T_{1}$, the nearest Mn spins are antiparallel in the $ab$ plane but parallel along $c$ axis., forming $C$-type AFM order with moment along $c$ axis. The magnetic space group is $P'/n'm'm$ (No. 129.416). Below $T_{2}$, there is an increase of (100)$\rm{_{T}}$, (101)$\rm{_{T}}$  and (102)$\rm{_{T}}$ peak intensities with ${\bf k}_{1}$ = (0,0,0)$\rm{_{T}}$ and simultaneously, new magnetic reflections with a propagation vector ${\bf k}_{2}$ = (0,0,1/2)$\rm{_{T}}$ appear. The comparison of the representative magnetic peak (0 0 1.5)$\rm{_{T}}$ below/above $T_{2}$ is shown in the inset of Fig. 3(a). Note that the 2nd weak peak with smaller omaga in the rocking curves is due to the presence of another tetragonal domain in the crystal. 

 \begin{figure} \centering \includegraphics [width = 1\linewidth] {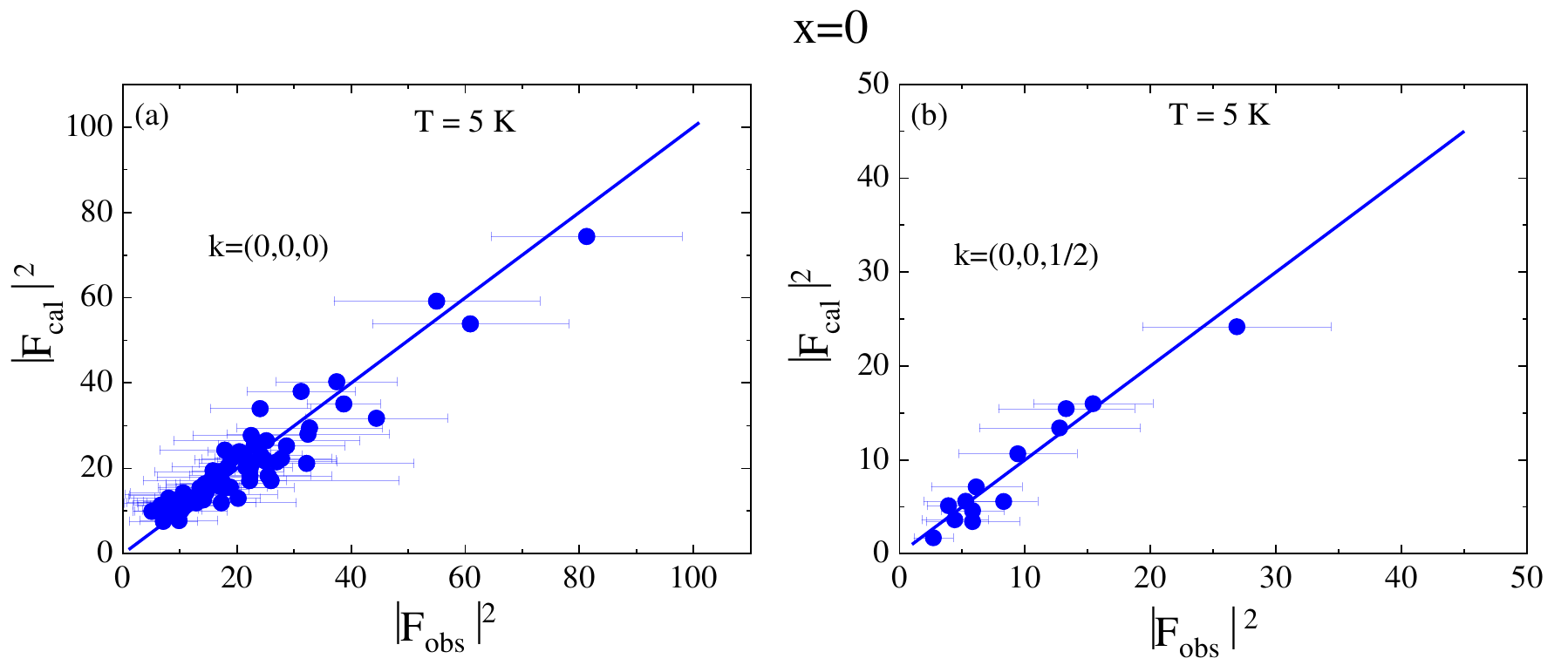}
\caption{(color online)  Comparison of the observed and calculated values of the squared
structure factors (a) for nuclear and magnetic peaks with {\bf k}=(0,0,0)$\rm{_{T}}$ , and (b) pure magnetic
peaks with {\bf k}=(0,0,1/2)$\rm{_{T}}$  taken at 5 K for x=0}
\label{fig:structure} 
\end{figure}

In $T<T_{2}$, two models of magnetic structures may result in an increase of (100)$\rm{_{T}}$  peak intensity: A). the canted Eu spins; B). a change of the Mn magnetic structure or ordered moment of Mn spins. We are able to exclude the scenario of canted Eu spins safely due to the following reasons: 1). the canted Eu spins require two \textbf{k} vectors, i.e., (0,0,0)$\rm{_{T}}$ and (0,0,1/2)$\rm{_{T}}$ on the Eu sublattice only. Spin canting can only be induced by the Dzyaloshinskii-Moriya (DM) interaction, but the
required DM vector is disallowed by lattice symmetry. Thus, such model is not symmetry-allowed magnetic structure of Eu sublattice. 2). The most likely model is the combination of $\Gamma_{2}$ for $\textbf{k}_{1} =(0,0,0)\rm{_{T}}$ and $\Gamma_{9}$ for ${\bf k}_{2}$ = (0,0,1/2)$\rm{_{T}}$, similar to the Eu canted order in $T_{3}<T<T_{2}$ in x=0.2 or 0.5 as shown below. However, adding such canted Eu model on C-type Mn order decreases the magnetic peak intensities of (101)$\rm{_{T}}$/(011)$\rm{_{T}}$ and (102)$\rm{_{T}}$/(012)$\rm{_{T}}$ and also leads to a rapid increase of (111)$\rm{_{T}}$ and (103)$\rm{_{T}}$ peak below $T_{2}$. However, both are inconsistent with the neutron results in EuMnSb$_{2}$. All the (101)$\rm{_{T}}$ (see Fig. S2(a)), (011)$\rm{_{T}}$, (102)$\rm{_{T}}$, and (012)$\rm{_{T}}$ increase obviously below $T<T_{2}$. Furthermore, there is no rapid increase of (111)$\rm{_{T}}$ or (113)$\rm{_{T}}$ (see Fig. S2(b)) peak within experimental uncertainty. Interestingly, as will be discussed below, such Eu canting model is consistent with the observation in $T_{3}<T<T_{2}$ for x=0.2 (see Fig. S2(c-d)) and x=0.5 compounds. 3). The fits to the neutron results using any of Eu canted models do not work and their refinement quality is much worse than that using the model B). Therefore, below $T_{2}$, while Eu spins forms a `` + + - -" order with moment along in-plane $a$  direction, the Mn spins keep $C$-type AFM order with an increased moment due to the Eu-Mn coupling to contribute to the increase of (100)$\rm{_{T}}$, (101)$\rm{_{T}}$ and (102)$\rm{_{T}}$ magnetic peak intensities. The magnetic structures are displayed in the right panel of Fig. 3 (d). The magnetic space group changes to Pn'm'a' (No. 62.449) in $T<T_{2}$. The refined magnetic moments for Eu and Mn spins at 5 K are 5.25(43)  and 4.63 (21) $\mu_{B}$, respectively. The comparisons between observed and calculated values of the squared structure factor for both  $\textbf{k}_{1} =(0,0,0)_{T}$ and $\textbf{k}_{2} =$ (0,0,1/2)$\rm{_{T}}$ are shown in Fig. S3 (a) and (b), respectively. We find a R$_{F}$ of 8.75 \% and $\chi^{2}$ of 0.28 for fits to $\textbf{k}_{1} =(0,0,0)_{T}$ dataset and R$_{F}$ of 8.93 \% and $\chi^{2}$ of 0.26 for $\textbf{k}_{2} =$ (0,0,1/2)$\rm{_{T}}$ dataset.

\textbf{2). x=0.2 }In the investigated temperature regions of Eu$_{1-x}$Sr$_{x}$MnSb$_{2}$ (x=0.2, 0.5, 0.8), all of the magnetic reflections can be indexed with the $Pnma$ unit cell with a magnetic propagation vector $\textbf{k} =$ (0,0,0)$\rm{_{O}}$. Note that the propagation vector of $\textbf{k} =$ (0,0,1/2)$\rm{_{T}}$ magnetic peaks of Eu sublattice needs to be updated to be  $\textbf{k} =$ (0,0,0)$\rm{_{O}}$ in orthorhombic notation in $T<T_{2}$ since the unit cell along out-of-plane direction in Eu$_{1-x}$Sr$_{x}$MnSb$_{2}$ (x=0.2, 0.5, 0.8) is doubled to that of EuMnSb$_{2}$. The symmetry of the allowed magnetic structures of $\textbf{k} =$ (0,0,0)$\rm{_{O}}$ determined from the SARAH program for Mn and Eu sublattices are summarized in Table \ref{basis_vector_table_3}. 

 \begin{figure} \centering \includegraphics [width = 1\linewidth] {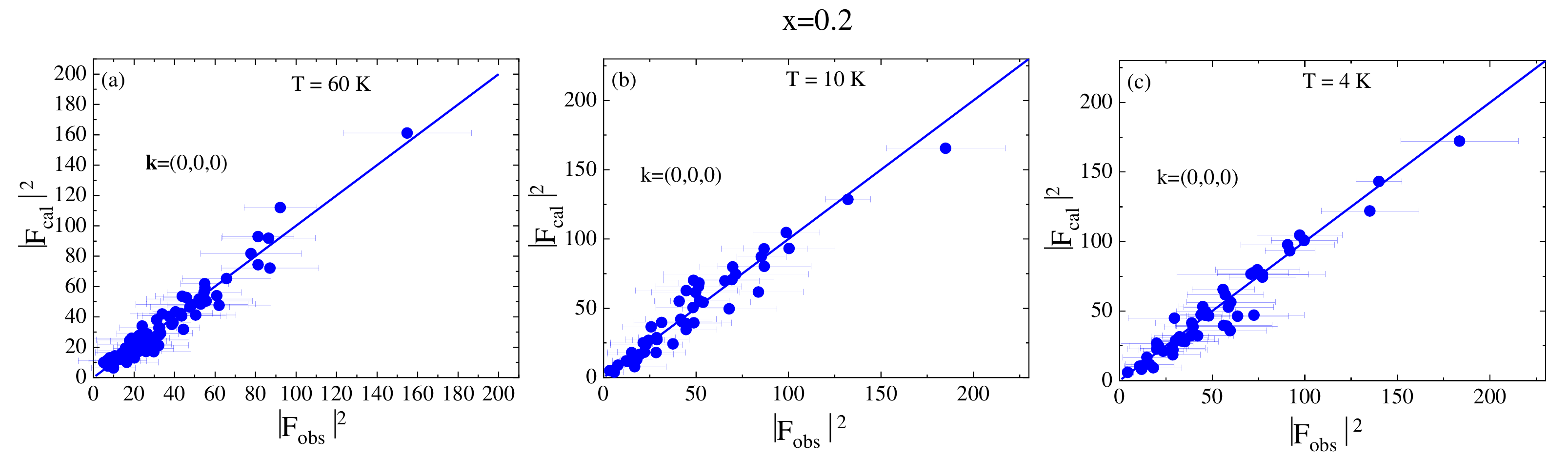}
\caption{(color online)  Comparison of the observed and calculated values of the squared structure factors for nuclear 
and magnetic peaks with {\bf k}=(0,0,0)$\rm{_{O}}$ for x=0.2 taken at (a) 60 K, (b) 10 K and (c) 4 K.}
\label{fig:structure} 
\end{figure}
 In $T_{2}<T<T_{1}$ for x=0.2, a C-type Mn order with moment along out-of-plane $a$ axis is also found, similar to that in $T_{2}<T<T_{1}$ of EuMnSb$_{2}$. The comparison of the observed and calculated square structure factors at 60 K is shown in Fig. S4 (a). In $T_{3}<T<T_{2}$, pure (H,K,L) (H=odd number) magnetic peaks appeared, corresponding to the $\textbf{k} =$ (0,0,1/2)$\rm{_{T}}$ magnetic peaks as in EuMnSb$_{2}$. Nevertheless, the simultaneous increase of (010)$\rm{_{O}}$ peak intensity below $T_{2}$ is much larger than that in EuMnSb$_{2}$ and the (010)$\rm{_{O}}$ peak intensity becomes almost three times of that observed at around 100 K, which implies a significant contribution from Eu spins. Furthermore, while (010)$\rm{_{O}}$, (211)$\rm{_{O}}$ and (611)$\rm{_{O}}$ peak intensities increase rapidly, the peak intensity of (210)$\rm{_{O}}$ and (410)$\rm{_{O}}$ decrease, as shown in Fig. S2(c-d).  The fits to the integration of nuclear and magnetic peaks at 10 K as shown in Fig. S4 (b) proves that a single representative analysis $\Gamma_{2}$ for both Mn and Eu sublattices is the correct one, i.e., the coexistence of C-type AFM Mn order with moment along $a$ axis and the canted Eu order in $ac$ plane (see the middle panel of Fig. 3 (e)). The  in-plane ``+ + - -" component of the Eu spins leads to magnetic symmetry breaking along the $a$ axis and therefore the appearance of magnetic peaks (H,K,L) (H=odd integers). The out-of-plane component has the spin configuration of ''+ - + -" , which decreases magnetic peaks like (210)$\rm{_{O}}$ (see Fig. S2 (c)) but increases the intensities of a few magnetic peaks such as (010)$\rm{_{O}}$, (211)$\rm{_{O}}$ and (611)$\rm{_{O}}$  (see Fig. S2 (d)). The magnetic space group for such magnetic structure is Pn'm'a' (No. 62.449). The refined magnetic moments of Mn and Eu at 10 K are 3.52(34) and 5.38(34) $\mu$ B, respectively, yielding a canting angle between Eu and Mn moments    $\approx 41 ^{\circ}$.

Upon cooling below $T<T_{3}$, the (300)$\rm{_{O}}$ magnetic peak intensity decreases, and simultaneously the (600)$\rm{_{O}}$ peak intensity increases due to the appearance of new magnetic peak (600)$\rm{_{O}}$ (see Fig. 3 (b)). This strongly indicates a spin-reorientation of Eu spins. The integrated intensities of nuclear and magnetic peaks can be best fitted (see Fig. S4 (c)) using the combination of $\Gamma_{2}$ and $\Gamma_{6}$. The obtained magnetic structure is displayed in the right panel of Fig. 3 (e), i.e., C-type AFM order of Mn spins with moment along $a$ axis and "`+ - + -`` A-type Eu order with canted moment in $ac$ plane. The refined Mn and Eu moments at 5 K are similar to those at 10 K, as expected for a spin reorientation transition. As shown in Fig. 3(b), there is a tendency that (300)$\rm{_{O}}$ peak intensity disappears and (600)$\rm{_{O}}$ peak intensity further increases while (010)$\rm{_{O}}$ does not show much change as the temperature approaches to 0 K. However, such magnetic structure remains valid as the ground state even at 0 K and the Eu canting to $a$ axis is necessary at 0 K since the Eu moment constricted to $c$ axis cannot account for the neutron results, for example very strong (010)$\rm{_{O}}$ magnetic peak.

 \begin{figure} \centering \includegraphics [width = 1\linewidth] {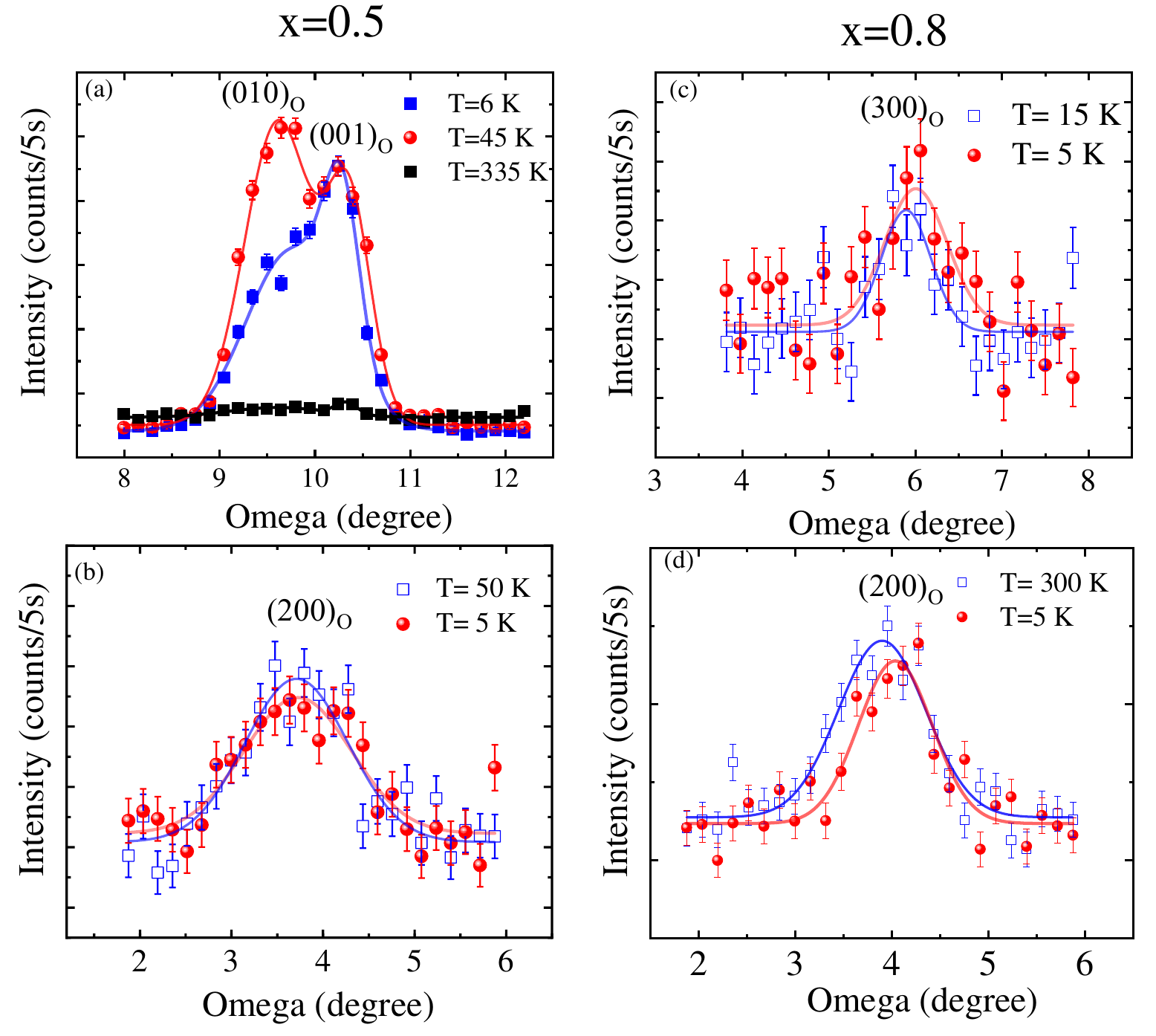}
\caption{(color online)  Rocking curves of (a) (010)$\rm{_{O}}$  and twinned (100)$\rm{_{O}}$, (b) (200)$\rm{_{O}}$ peaks for x=0.5 at different temperatures. Rocking curves of (c) (300)$\rm{_{O}}$  and (d) (200)$\rm{_{O}}$ peaks at different temperatures for x=0.8.}
\label{fig:structure} 
\end{figure}

\textbf{3). x=0.5 }In $T_{2}<T<T_{1}$ of x=0.5, a C-type Mn AFM order with moment along $a$ axis is also found by fitting to the neutron results, with ordered moment $\approx$ 3.74(15) $\mu$ B at 50 K. Below $T_{2}$, (H,K,L) (H=odd number) magnetic peaks appeared, and the intensity of (010)$\rm{_{O}}$ (see Fig. 3(c) and S5 (a)) and (211)$\rm{_{O}}$ magnetic peak also increase. However, there is no clear increase of (200)$\rm{_{O}}$ (see Fig. S5(b)) and (600)$\rm{_{O}}$ peak intensities. All of these observations are very similar to that in  $T_{3}<T<T_{2}$ of x=0.2. Fits to the integration of nuclear and magnetic peaks at 5 K shown in Fig. S6 (a) confirm the similar magnetic structure to that in $T_{3}<T<T_{2}$ of x=0.2, as also shown in the middle panel of Fig. 3 (e). Compared to x=0.2, the main difference is that the canting angle of Eu spins with respect to $a$ axis becomes smaller, $\approx 24 ^{\circ}$ at 5 K. 
 \begin{figure} \centering \includegraphics [width = 1\linewidth] {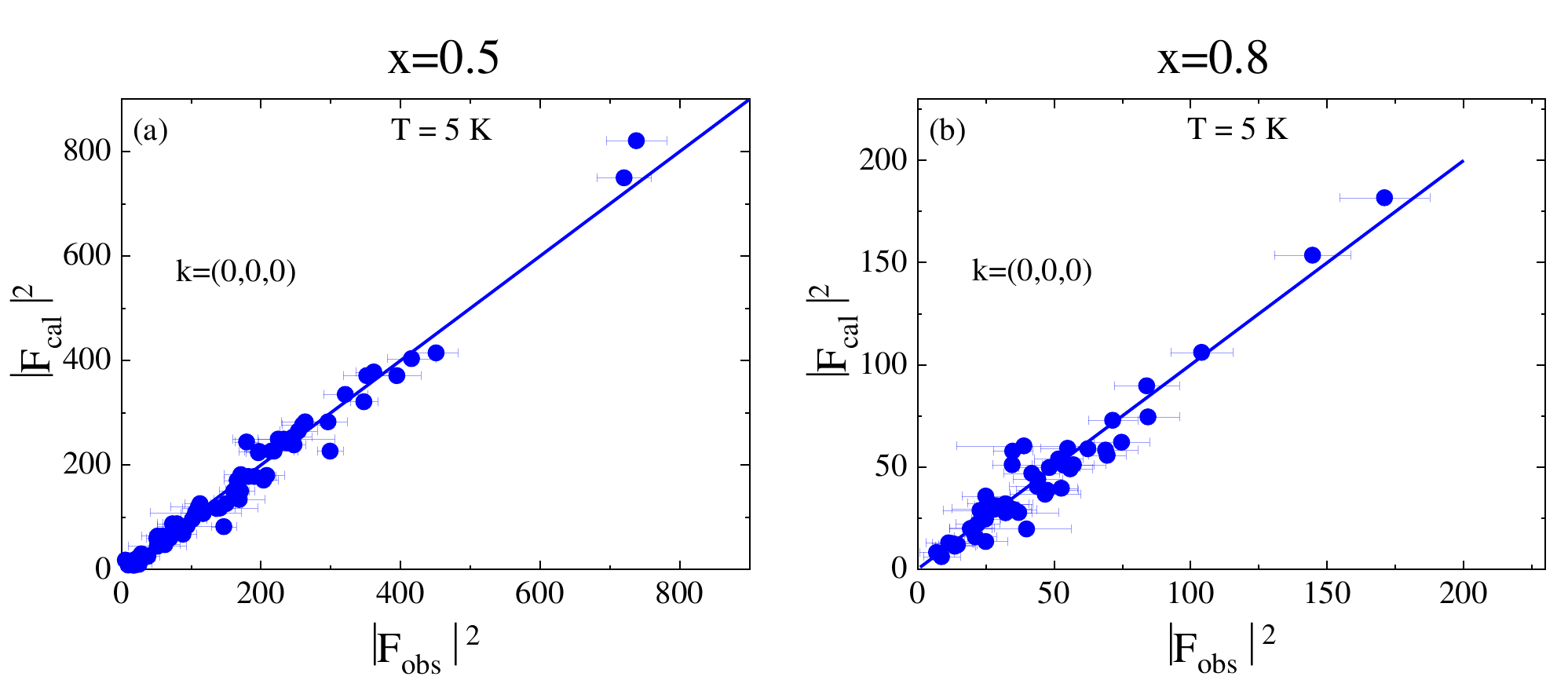}
\caption{(color online)  Comparison of the observed and calculated values of the
squared structure factors for nuclear and magnetic peaks with {\bf k}=(0,0,0)$\rm{_{O}}$ taken at 5 K for
(a) x= 0.5 and (b) x=0.8 compounds.}
\label{fig:structure} 
\end{figure}

\textbf{4). x=0.8 }As the doping of Sr increases to 0.8 for x=0.8, a C-type AFM Mn order is also determined in $T_{2}<T<T_{1}$ (see the left panel of Fig. 3(f)). Very interesting, below $T_{2}$, no (H,K,L) (H=odd number) magnetic peaks appeared (see Fig. 3(c) and S5 (c)), indicative of the absence of the in-plane ``+ + - -'' Eu component. There are no (200) or (600) magnetic peaks either (see Fig. S5 (d)), suggesting there is no in-plane ``+ - + -'' Eu component. The increase of the (010)$\rm{_{O}}$ magnetic peak below $T_{2}$ suggested there is still out-of-plane ``+ - + -" Eu component along the $a$ axis. Indeed, the integration of the nuclear and magnetic peaks at 5 K can be best fitted (see Fig. S 6(b)) using a single representative analysis $\Gamma_{2}$ with both Mn and Eu moments pointing along the same $a$ axis, with the magnetic space group Pn'm'a' (No. 62.449). At 5 K, the moments of Mn and Eu are refined to be 3.80(22) $\mu$ B and 5.17(58) $\mu$ B, respectively. The magnetic structure is illustrated in the right panel of Fig. 3(f).

\textbf{Extraction of nontrivial Berry phase for x=0.5 and 0.8\\}
Figure S7 (a) displays the Landau level (LL) fan diagram established on the base of oscillatory resistivity $\rho_{in}$ for $x$=0.5. From the extrapolation of the best linear fit to the fan diagram, we obtained the intercept n$_{0}$ on the n axis to be 0.38. The corresponding Berry phase is 2$\pi \times$ n$_{0}$=0.76 $\pi$, close to the non-trivial $\pi$ Berry phase. For $x$=0.8, the fits to LL fan diagram based on the oscillatory resistivity $\rho_{in}$ and $\rho_{out}$  are shown in the Fig. S 7 (b) and the inset of Fig. 4 (f), respectively. We obtained intercept n$_{0}$ of 0.44 and 0.57, and the corresponding Berry phases of 0.88 $\pi$ and 1.14 $\pi$, respectively. Both are also close to non-trivial $\pi$ Berry phase.  

 \begin{figure} \centering \includegraphics [width = 1\linewidth] {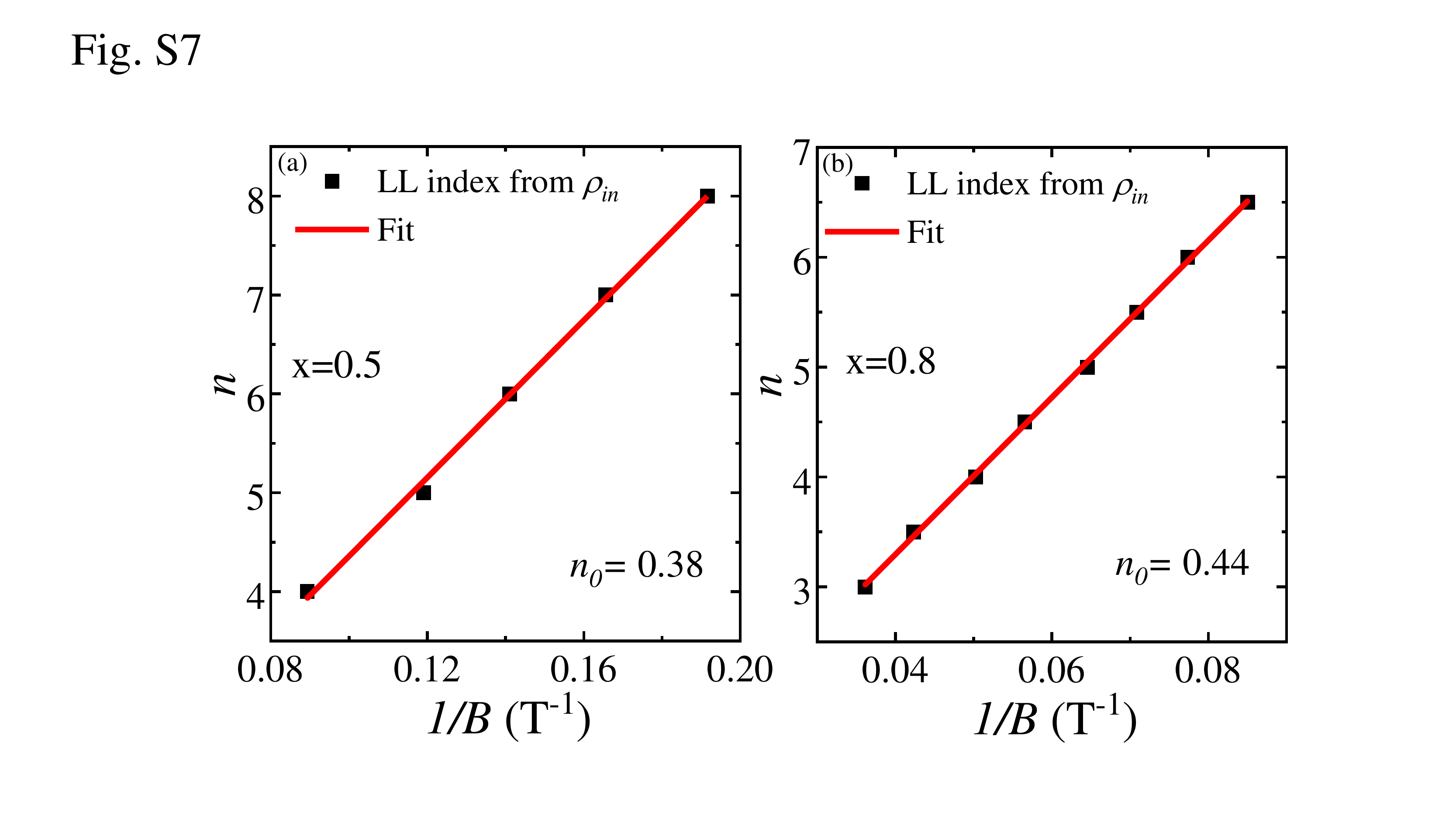}
\caption{(color online)  Linear fits to the Landau Level (LL) index fan diagram at 1.8 K
derived from the oscillatory resistivity $\rho_{in}$ for (a) x=0.5 and (b) x=0.8.}
\label{fig:structure} 
\end{figure}

 \begin{table} 
 \centering
  \setlength{\abovecaptionskip}{0pt}%
\setlength{\belowcaptionskip}{10pt}%
 \caption{Crystallographic parameters and residuals for EuMn$_{1-x}$Sb$_{2}$ models fit to single crystal x-ray diffraction data on x=0 compound at $T =$ 293 K.  The statistical uncertainties are given in parentheses.}
\renewcommand{\arraystretch}{0.8}
 \begin{tabular}{ cccccccccccccccccccc|cccccccccccccccccccc }
\hline\hline

\multicolumn{20}{c}{Formula}  &  \multicolumn{20}{c}{ EuMn$_{0.912(2)}$Sb$_{2}$ }    \\
  \hline
\multicolumn{20}{c}{Crystal System}  &  \multicolumn{20}{c}{ Tetragonal }    \\
 \multicolumn{20}{c}{Space Group}  &  \multicolumn{20}{c}{ $P4/nmm$}    \\
\multicolumn{20}{c}{$a$ ($\AA{}$) }  &  \multicolumn{20}{c}{ 4.31246(3) }    \\
\multicolumn{20}{c}{$c$ ($\AA{}$) }  &  \multicolumn{20}{c}{ 11.2763(8) }    \\
\multicolumn{20}{c}{$V$ ($\AA{}^{3}$)}  &  \multicolumn{20}{c}{ 216.78(3) }    \\
\multicolumn{20}{c}{$Z$)}  &  \multicolumn{20}{c}{ 2 }    \\
\multicolumn{20}{c}{Goodness of fit}  &  \multicolumn{20}{c}{ 1.24}    \\
\multicolumn{20}{c}{$R_{1}[F^{2}>2 \sigma(F^{2})]^{a})$}  &  \multicolumn{20}{c}{0.053}    \\
\multicolumn{20}{c}{$wR_{2}(F^{2})^{b}$}  &  \multicolumn{20}{c}{0.145}    \\

   \hline
$   ^{a}R_{1}(F)=\sum  ||F_{o}|-|F_{c}||/\sum  |F_{o}|  $ \\
$ ^{b}wR_{2}(F^{2})=[\sum [w (F_{o}^{2}-F_{c}^{2})^{2}]/\sum [w (F_{o}^{2})^{2}]^{1/2} $

\end{tabular}

\label{crystal_1}
\end{table}

\begin{table} 
\centering
\setlength{\abovecaptionskip}{0pt}%
\setlength{\belowcaptionskip}{10pt}%
\caption{Atomic fractional coordinates and U$_{equs}$ for EuMnSb$_{2}$ obtained from single crystal x-ray data at 293 K.  $ ^{a}U_{eq}$ is defined as 1/3 of the trace of the orthogonalized $U_{ij }$ tensor.}
\renewcommand{\arraystretch}{0.8}
\begin{tabular}{c|c|c|c|c|c}
\hline\hline
 atom & Wyckoff Site&  x & y &  z& $ ^{a}U_{eq}$    \\
 \hline
Eu & 2c &    0.25&  0.25     &   0.7271(1)  &  0.0151(5)   \\
Mn & 2a &    0.75&  0.25     &   0   &  0.0179(5)  \\ 
Sb1 & 2b &    0.75&  0.25     &   0.5  &  0.0176(5)   \\
 Sb2 & 2c &    0.25&  0.25     &   0.01510(2)  &  0.0166(6)   \\
 \hline\hline  

\end{tabular}

\label{crystal_2}
\end{table}

\begin{table} 
\centering
\small
\setlength{\tabcolsep}{6pt}
\renewcommand{\arraystretch}{0.8}
\setlength{\abovecaptionskip}{0pt}%
\setlength{\belowcaptionskip}{10pt}%
\caption{Basis vectors for the space group $P 4/n m m$ with 
${\bf k}=( 0,~ 0,~ 0)\rm{_{T}}$ in EuMnSb$_{2}$. The Mn atoms of the nonprimitive 
basis are defined according to 
Mn$_{1}$ : $( .75,~ .25,~ 0)$, Mn$_{2}$: $( .25,~ .75,~ 0)$.  The Eu atoms of the nonprimitive 
basis are defined according to 
1: $( .25,~ .25,~ .729)$, 2: $( .75,~ .75,~ .271)$ at 5 K.} 
\renewcommand{\arraystretch}{0.8}
\begin{tabular}{ccc|cccccc}
\hline\hline
  IR  &  BV  &  Atom & \multicolumn{6}{c}{BV components}\\
      &      &             &$m_{\|a}$ & $m_{\|b}$ & $m_{\|c}$ &$im_{\|a}$ & $im_{\|b}$ & $im_{\|c}$ \\
\hline
$\Gamma_{3}$ & $ \psi_{1}$ &      Mn$_{1}$ &      0 &      0 &      8 &      0 &      0 &      0  \\
             &              &      Mn$_{2}$ &      0 &      0 &      8 &      0 &      0 &      0  \\
$\Gamma_{6}$ & $\psi_{2}$ &      Mn$_{1}$ &      0 &      0 &      8 &      0 &      0 &      0  \\
             &              &      Mn$_{2}$ &      0 &      0 &     -8 &      0 &      0 &      0  \\
$\Gamma_{9}$ & $\psi_{3}$ &      Mn$_{1}$ &      4 &      0 &      0 &      0 &      0 &      0  \\
             &              &      Mn$_{2}$ &      4 &      0 &      0 &      0 &      0 &      0  \\
             & $\psi_{4}$ &      Mn$_{1}$ &      0 &     -4 &      0 &      0 &      0 &      0  \\
             &              &      Mn$_{2}$ &      0 &     -4 &      0 &      0 &      0 &      0  \\
$\Gamma_{10}$ & $\psi_{5}$ &      Mn$_{1}$ &      0 &      4 &      0 &      0 &      0 &      0  \\
             &              &      Mn$_{2}$ &      0 &     -4 &      0 &      0 &      0 &      0  \\
             & $\psi_{6}$ &      Mn$_{1}$ &     -4 &      0 &      0 &      0 &      0 &      0  \\
             &              &      Mn$_{2}$ &      4 &      0 &      0 &      0 &      0 &      0  \\
\hline
$\Gamma_{2}$ & $\psi_{1}$ &      Eu$_{1}$ &      0 &      0 &      8 &      0 &      0 &      0  \\
             &              &      Eu$_{2}$ &      0 &      0 &     -8 &      0 &      0 &      0  \\
$\Gamma_{3}$ & $\psi_{2}$ &       Eu$_{1}$ &      0 &      0 &      8 &      0 &      0 &      0  \\
             &              &       Eu$_{2}$ &      0 &      0 &      8 &      0 &      0 &      0  \\
$\Gamma_{9}$ & $\psi_{3}$ &       Eu$_{1}$ &      4 &      0 &      0 &      0 &      0 &      0  \\
             &              &       Eu$_{2}$ &      4 &      0 &      0 &      0 &      0 &      0  \\
             & $\psi_{4}$ &       Eu$_{1}$ &      0 &     -4 &      0 &      0 &      0 &      0  \\
             &              &       Eu$_{2}$ &      0 &     -4 &      0 &      0 &      0 &      0  \\
$\Gamma_{10}$ & $\psi_{5}$ &       Eu$_{1}$ &      0 &      4 &      0 &      0 &      0 &      0  \\
             &              &       Eu$_{2}$ &      0 &     -4 &      0 &      0 &      0 &      0  \\
             & $\psi_{6}$ &       Eu$_{1}$ &      4 &      0 &      0 &      0 &      0 &      0  \\
             &              &       Eu$_{2}$ &     -4 &      0 &      0 &      0 &      0 &      0  \\
\hline\hline
\end{tabular}

\label{basis_vector_table_1}
\end{table}

\begin{table} 
\centering
\setlength{\abovecaptionskip}{0pt}%
\setlength{\belowcaptionskip}{10pt}%
\caption{Basis vectors for the space group $P 4/n m m$ with 
${\bf k}=( 0,~ 0,~ 1/2)\rm{_{T}}$ in EuMnSb$_{2}$. The Eu atoms of the nonprimitive 
basis are defined according to 
1: $( .25,~ .25,~ .729)$, 2: $( .75,~ .75,~ .271)$ at 5 K.  }

\renewcommand{\arraystretch}{0.8}
\begin{tabular}{ccc|cccccc}
\hline\hline
  IR  &  BV  &  Atom & \multicolumn{6}{c}{BV components}\\
      &      &             &$m_{\|a}$ & $m_{\|b}$ & $m_{\|c}$ &$im_{\|a}$ & $im_{\|b}$ & $im_{\|c}$ \\
\hline
$\Gamma_{2}$ & $\psi_{1}$ &      Eu$_{1}$ &      0 &      0 &      8 &      0 &      0 &      0  \\
             &              &      Eu$_{2}$ &      0 &      0 &      8 &      0 &      0 &      0  \\
$\Gamma_{3}$ & $\psi_{2}$ &      Eu$_{1}$ &      0 &      0 &      8 &      0 &      0 &      0  \\
             &              &      Eu$_{2}$ &      0 &      0 &     -8 &      0 &      0 &      0  \\
$\Gamma_{9}$ & $\psi_{3}$ &      Eu$_{1}$ &      4 &      0 &      0 &      0 &      0 &      0  \\
             &              &      Eu$_{2}$ &     -4 &      0 &      0 &      0 &      0 &      0  \\
             & $\psi_{4}$ &      Eu$_{1}$ &      0 &     -4 &      0 &      0 &      0 &      0  \\
             &              &      Eu$_{2}$ &      0 &      4 &      0 &      0 &      0 &      0  \\
$\Gamma_{10}$ & $\psi_{5}$ &      Eu$_{1}$ &      0 &      4 &      0 &      0 &      0 &      0  \\
             &              &      Eu$_{2}$ &      0 &      4 &      0 &      0 &      0 &      0  \\
             & $\psi_{6}$ &      Eu$_{1}$ &      4 &      0 &      0 &      0 &      0 &      0  \\
             &              &      Eu$_{2}$ &      4 &      0 &      0 &      0 &      0 &      0  \\
             \hline\hline
\end{tabular}

\label{basis_vector_table_2}
\end{table}

\begingroup

\begin{table} 
\centering
\small
\setlength{\tabcolsep}{6pt}
\renewcommand{\arraystretch}{0.44}
\setlength{\abovecaptionskip}{0pt}%
\setlength{\belowcaptionskip}{10pt}%
\caption{Basis vectors for the space group $P n m a$ with 
${\bf k}=( 0,~ 0,~ 0)\rm{_{O}}$. Note that Mn and Eu occupy the same Wyckoff site, as listed in Table S3.}
\begin{tabular}{ccc|cccccc}
\hline\hline
  IR  &  BV  &  Atom & \multicolumn{6}{c}{BV components}\\
      &      &             &$m_{\|a}$ & $m_{\|b}$ & $m_{\|c}$ &$im_{\|a}$ & $im_{\|b}$ & $im_{\|c}$ \\
\hline
$\Gamma_{1}$ & $\psi_{1}$  &     1 &      0 &      2 &      0 &      0 &      0 &      0  \\
             &              &     2 &      0 &     -2 &      0 &      0 &      0 &      0  \\
             &              &     3 &      0 &      2 &      0 &      0 &      0 &      0  \\
             &              &     4 &      0 &     -2 &      0 &      0 &      0 &      0  \\
$\Gamma_{2}$ & $\psi_{2}$ &     1 &      2 &      0 &      0 &      0 &      0 &      0  \\
             &              &     2 &      2 &      0 &      0 &      0 &      0 &      0  \\
             &              &     3 &     -2 &      0 &      0 &      0 &      0 &      0  \\
             &              &     4 &     -2 &      0 &      0 &      0 &      0 &      0  \\
             & $\psi_{3}$ &     1 &      0 &      0 &      2 &      0 &      0 &      0  \\
             &              &     2 &      0 &      0 &     -2 &      0 &      0 &      0  \\
             &              &     3 &      0 &      0 &     -2 &      0 &      0 &      0  \\
             &              &     4 &      0 &      0 &      2 &      0 &      0 &      0  \\
$\Gamma_{3}$ & $\psi_{4}$ &     1 &      2 &      0 &      0 &      0 &      0 &      0  \\
             &              &     2 &      2 &      0 &      0 &      0 &      0 &      0  \\
             &              &     3 &      2 &      0 &      0 &      0 &      0 &      0  \\
             &              &     4 &      2 &      0 &      0 &      0 &      0 &      0  \\
             & $\psi_{5}$ &     1 &      0 &      0 &      2 &      0 &      0 &      0  \\
             &              &     2 &      0 &      0 &     -2 &      0 &      0 &      0  \\
             &              &     3 &      0 &      0 &      2 &      0 &      0 &      0  \\
             &              &     4 &      0 &      0 &     -2 &      0 &      0 &      0  \\
$\Gamma_{4}$ & $\psi_{6}$ &     1 &      0 &      2 &      0 &      0 &      0 &      0  \\
             &              &     2 &      0 &     -2 &      0 &      0 &      0 &      0  \\
             &              &     3 &      0 &     -2 &      0 &      0 &      0 &      0  \\
             &              &     4 &      0 &      2 &      0 &      0 &      0 &      0  \\
$\Gamma_{5}$ & $\psi_{7}$ &     1 &      0 &      2 &      0 &      0 &      0 &      0  \\
             &              &     2 &      0 &      2 &      0 &      0 &      0 &      0  \\
             &              &     3 &      0 &      2 &      0 &      0 &      0 &      0  \\
             &              &     4 &      0 &      2 &      0 &      0 &      0 &      0  \\
$\Gamma_{6}$ & $\psi_{8}$ &     1 &      2 &      0 &      0 &      0 &      0 &      0  \\
             &              &     2 &     -2 &      0 &      0 &      0 &      0 &      0  \\
             &              &     3 &     -2 &      0 &      0 &      0 &      0 &      0  \\
             &              &     4 &      2 &      0 &      0 &      0 &      0 &      0  \\
             & $\psi_{9}$ &     1 &      0 &      0 &      2 &      0 &      0 &      0  \\
             &              &     2 &      0 &      0 &      2 &      0 &      0 &      0  \\
             &              &     3 &      0 &      0 &     -2 &      0 &      0 &      0  \\
             &              &     4 &      0 &      0 &     -2 &      0 &      0 &      0  \\
$\Gamma_{7}$ & $\psi_{10}$ &     1 &      2 &      0 &      0 &      0 &      0 &      0  \\
             &              &     2 &     -2 &      0 &      0 &      0 &      0 &      0  \\
             &              &     3 &      2 &      0 &      0 &      0 &      0 &      0  \\
             &              &     4 &     -2 &      0 &      0 &      0 &      0 &      0  \\
             & $\psi_{11}$ &     1 &      0 &      0 &      2 &      0 &      0 &      0  \\
             &              &     2 &      0 &      0 &      2 &      0 &      0 &      0  \\
             &              &     3 &      0 &      0 &      2 &      0 &      0 &      0  \\
             &              &     4 &      0 &      0 &      2 &      0 &      0 &      0  \\
$\Gamma_{8}$ & $\psi_{12}$ &     1 &      0 &      2 &      0 &      0 &      0 &      0  \\
             &              &     2 &      0 &      2 &      0 &      0 &      0 &      0  \\
             &              &     3 &      0 &     -2 &      0 &      0 &      0 &      0  \\
             &              &     4 &      0 &     -2 &      0 &      0 &      0 &      0  \\
             \hline\hline
             \end{tabular}

\label{basis_vector_table_3}
\end{table}
\endgroup

\end{document}